\documentclass[%
 reprint,
 amsmath,amssymb,
 aps,
floatfix,
]{revtex4-2}

\usepackage{siunitx}
\usepackage{graphicx}% Include figure files
\usepackage{dcolumn}% Align table columns on decimal point
\usepackage{bm}% bold math
\usepackage{physics}
\usepackage{braket}
\usepackage{dsfont}
\usepackage{hyperref}
\usepackage{commath}
\hypersetup{
anchorcolor=blue,
colorlinks=true,
citecolor=blue,
urlcolor=blue,
linkcolor=blue}
\usepackage{cleveref}
\usepackage{float}

\renewcommand{\ketbra}[2]{\ensuremath{\ket{#1}\hspace{-1pt}\bra{#2}}}

\newcommand{\statevariable}{X}
\newcommand{\meas}{\Pi} % measurement operator
\newcommand{\vecrho}{\vec\rho}
\newcommand{\vecp}{\vec b}
\newcommand{\imi}{\mathrm{i}}
\newcommand{\expo}[1]{\mathrm{e}^{#1}}
\newcommand{\expup}[1]{\mathrm{e}^{#1}}
\newcommand{\rhotherm}[1]{\ensuremath{\rho^\mathrm{th}_{#1}}}

\newcommand{\ptherm}[1]{\ensuremath{P^\mathrm{th}_{#1}}}
\newcommand{\xmax}{x_\text{max}}
\newcommand{\pmax}{p_\text{max}}
\newcommand{\amax}{\alpha_\text{max}}
\newcommand{\nthermal}{n_\text{th}}

\newcommand{\rhoideal}{\rho_\text{ideal}}
\newcommand{\sparam}[2]{\tilde{W}(#1,#2)}

\newcommand{\figpanel}[2]{\hyperref[#1]{\ref*{#1}(#2)}}

\begin{document}

\title{Simple, reliable and noise-resilient continuous-variable quantum state tomography with convex optimization}% Force line breaks with \\

\author{Ingrid Strandberg}
%\email[e-mail:]{email1}
%\author{Shahnawaz Ahmed}
%\author{Fernando Quijandr\'{\i}a}

\affiliation{Department of Microtechnology and Nanoscience MC2, Chalmers University of Technology, SE-412 96
G\"oteborg, Sweden}

\date{\today}% It is always \today, today,
             %  but any date may be explicitly specified

\begin{abstract}

Precise reconstruction of unknown quantum states from measurement data, a process commonly called quantum state tomography, is a crucial component in the development of quantum information processing technologies.
Many different tomography methods have been proposed over the years. Maximum likelihood estimation is a prominent example, being the most popular method for a long period of time. Recently, more advanced neural network methods have started to emerge. Here, we go back to basics and present a method for continuous variable state reconstruction that is both conceptually and practically simple, based on convex optimization. Convex optimization has been used for process tomography and qubit state tomography, but seems to have been overlooked for continuous variable quantum state tomography. We demonstrate high-fidelity reconstruction of an underlying state from data corrupted by thermal noise and imperfect detection, for both homodyne and heterodyne measurements. A major advantage over other methods is that convex optimization algorithms are guaranteed to converge to the optimal solution.

%One of the advantages of LRE is that the MSE upper bound can be given analytically as , which is dependant explicitly upon the measurement bases. 
% https://www.nature.com/articles/srep03496

%keyworks: quantum state reconstruction, quantum state estimation, quantum state tomography, quantum optics, convex optimization, homodyne detection, heterodyne detection, Wigner function, quantum information, continuous variable quantum computing

%Quantum state tomography is the process by which an unkown quantum state is completely characterized by measurement data

\end{abstract}

%\keywords{Suggested keywords}%Use showkeys class option if keyword
                              %display desired
\maketitle

%\tableofcontents

\section{Introduction}

%\textcolor{red}{show that it works for more states. GKP for example. appendix}
%\textcolor{red}{table fidelities cat with different alphas}
%\textcolor{red}{implement noise compensated homodyne}
%\textcolor{red}{fast, reliable, transparent}
%\textcolor{red}{plot fidelity vs nr of gridpoints}

Quantum state tomography has a rich history, and interest in it is only increasing due to the current development of many quantum technologies such as quantum sensing, communication, cryptography and computing.
The goal of quantum state tomography is to infer information about a quantum state from measurement data.
Calculating probability distributions for observables given a specific quantum state is straightforward, but the inverse problem of calculating a quantum state given measured probability distributions is fraught with issues due to the problem being \emph{ill-posed}, meaning slight variations or noise in the measurement data can heavily affect the result~\cite{Kabanikhin2011Dec}.

While nowadays multi-qubit state tomography is an area of great interest, the development of quantum tomography started with continuous variable (CV) states in 1989 when Vogel and Risken proposed that the Wigner function can be obtained from the inverse Radon transform of homodyne measurement data~\cite{Vogel1989Sep}.
%Smithey \emph{et.\ al}.~\cite{Smithey1993Mar} were the first to do this with experimental data obtained by homodyne detection. They used this method for the reconstruction of a squeezed state in 1992, one of the earliest types of generated quantum states.
Ever since the first experimental quantum tomography was performed using such measurements~\cite{Smithey1993Mar}, homodyne tomography has been a mainstay in the quantum optics community. At the beginning, reconstruction techniques had the issue that the resulting states could be unphysical~\cite{Lvovsky2009Mar}.
%However, the inverse Radon transformation is severely ill-posed, rendering it numerically unstable. In Ref.~\cite{Smithey1993Mar} they mitigated this by using a filtered back-projection algorithm used for classical tomography in the medical field. Unfortunately, a problem with this approach is that it does not preserve positive semidefiniteness of the reconstructed state, meaning the result can be unphysical~\cite{Lvovsky2009Mar}.
%
%Additionally, while the density matrix and Wigner function of a state are one-to-one and calculating the Wigner function from a given state is straightforward, obtaining the density matrix from the Wigner function is also an inverse problem in itself. %Methods of obtaining the density matrix directly without the detour through the Wigner function were developed,for example using \textcolor{blue}{pattern functions}~\cite{Leonhardt1995Dec} whose analytical expressions can take complicated forms~\cite{Richter1996Mar} \textcolor{blue}{The pattern functions can be complicated to compute}.
New numerical methods were developed to handle quantum state reconstruction from homodyne measurements, ranging from pattern function methods in the mid 90s~\cite{Leonhardt1995Dec, Richter1996Mar} to maximum likelihood estimation (MLE) emerging in the late 90s~\cite{Hradil1997Mar, Banaszek1999Dec}. %Maximum likelihood estimation (MLE) is a statistical inference method known from classical statistics.
An iterative MLE algorithm was introduced in the 2000s~\cite{Lvovsky2004May}, and it has remained very popular over time, with different variations offering improvements being developed~\cite{Rehacek2007Apr, Shang2017Jun, Blume-Kohout2010Nov,Baumgratz2013Dec}.
%\textcolor{blue}{use for experimental data? check diluted ref 14}.
Currently, even more advanced methods are emerging. Bayesian methods have been proposed~\cite{Blume-Kohout2010Apr, Granade2016Mar,Lukens2020Jun}, %\textcolor{blue}{but these require the use of Monte Carlo sampling algorithms.}.
and neural networks for CV state tomography have started to appear~\cite{Tiunov2020May, Ghosh2020Jul,Ahmed2021Sep1,Ahmed2021Sep}. There are downsides to all above mentioned methods: Bayesian methods require the use of Monte Carlo sampling algorithms, iterations of an MLE algorithm have to be terminated at some arbitrary point, and neural networks are practically black boxes where the result can depend strongly on the chosen cost function.

%for the verification of state preparation, the analysis of quantum dynamics and decoherence, and for retrieving information encoded in quantum states.
Quantum state tomography is an essential tool in the field of quantum information. Not only is it needed to characterize states and verify state preparation, having a reliable method of state tomography also plays an important role in several methods for quantum process tomography, which characterizes a quantum gate or channel~\cite{Chuang1997Nov,Bialczak2010Jun,Altepeter2003May,Mohseni2008Mar,Rahimi-Keshari2011Jan, Ghalaii2017Mar}.

Here, we demonstrate a method for continuous variable state reconstruction that is both conceptually and practically simple, based on convex optimization.
Convex optimization comprise a special class of mathematical optimization problems that can be solved numerically very reliably and efficiently~\cite{Boyd2004Mar}. It has been applied to process tomography~\cite{Huang2020Feb, Kim2021Apr}, detector tomography~\cite{Zhang2012Jun, Cooper2014Jul} and discrete variable state-tomography (though often assuming low-rank density matrices)~\cite{Riofrio2017May, Flammia2012Sep,Gross2010Oct,Smith2013Mar}. Still, it seems to have been overlooked for CV state tomography. We use it to demonstrate high-fidelity reconstruction of a CV quantum state with no assumptions of the rank, from data contaminated by thermal noise or imperfect detection. The primary advantage of this method compared to iterative maximum likelihood and neural network methods is that convex optimization algorithms are guaranteed to converge to the optimal solution.

The paper is structured as follows. We begin by introducing the basic concepts of quantum states and measurements in Section~\ref{sec:tomography}, and how their properties are suitable for forming the tomography problem as a convex optimization problem. In Section~\ref{sec:application} our state tomography method is demonstrated first on simulated homodyne data and then heterodyne data. For heterodyne measurement, its performance is compared to the ubiquitous maximum likelihood estimation and a new neural network tomography method. We also test our method on real experimental data with good results.
Finally, we conclude and summarize in Section~\ref{sec:conclusions}.

%Ref.~\cite{Nehra2020Oct} uses it, but the point of the paper is to introduce what they call "Generalized overlap quantum state tomography", which uses a Wigner function overlap measurement rather than the Wigner function to reconstruct the density matrix. Our implementation requires no new type of measurement, and it is versatile in the sense that it can use either heterodyne, homodyne or Wigner measurements. A major advantage over other methods is that convex optimization algorithms are guaranteed to converge to the optimal solution.

%Here, we show a super simple and fast way to perform CV state tomography, utilizing convex optimization. Convex optimization has been applied to %process tomography~\cite{Huang2020Feb, Kim2021Apr}, detector tomography~\cite{Zhang2012Jun, Cooper2014Jul} and discrete variable state-tomography,
%(often assuming low-rank density matrices)~\cite{Riofrio2017May, Flammia2012Sep,Gross2010Oct,Smith2013Mar}.

%%%%%%%%%%%%%%%%%%%%%%%%%%%%%%%%%%%%%%%%%%%%%%%%%%%%%%%%%%%%%%%%%%%%%%%%%%%%%%%%%%%%%%%%%%%%%%%%%%%%%%%%%%%%%%%%%%%%%%%%%%%%
\section{Quantum state tomography}~\label{sec:tomography}

\subsection{The quantum state}

The state of a quantum system can be described by a $N\times N$ density matrix $\rho$ in a Hilbert space, where $N$ is the dimension of the system in question. For a CV state, $\rho$ is often presented in the Fock (number state) basis. In theory the corresponding state space is infinite dimensional, but for practical reasons one may truncate the Hilbert space to a finite dimension that is sufficient to contain all non-zero elements of the density matrix~\footnote{If one does not have a priori information about the maximum photon population, to ensure there is no error due to the cutoff one can increase it until the result does not change with increasing dimension.}.
There are a few conditions a density matrix must fulfill in order to describe a physical system: it must be Hermitian, have unit trace, and have no negative eigenvalues~\cite{Nielsen2010-fe}.
%-----------------------------------
\subsection{Measurements}

A general quantum measurement can be described by a set of operators $\{ \meas_k \}$ called \emph{positive operator valued measure} (POVM), where each operator is associated with a possible measurement outcome $k$~\footnote{In general, $k$ can be continuous. Nevertheless, one can always construct POVM elements labeled by a discrete index by binning the measurement results. This will be done for homodyne data in section~\ref{sec:homodyne}.} such that the probability of obtaining $k$ is~\cite{Peres2002}
\begin{equation}\label{eq:measurement_probability}
    p_k = \Tr[\meas_k \rho].
\end{equation}
When the density matrix $\rho$ is reshaped into a vector $\vecrho$ (see Appendix~\ref{app:vectorize}), the tomography problem can be expressed as a linear equation
\begin{equation}\label{eq:linear_eq}
    A\vecrho=\vecp,
\end{equation}
where the state $\vecrho$ is the unknown to be solved for. The elements $b_k$ of $\vecp$ are the measurement records $p_k$~\footnote{The notation $\vecp$ is chosen instead of $\vec p$ to avoid confusion with the vectorized density matrix $\vec \rho$}.%~\eqref{eq:measurement_probability}.
The matrix $A$ contains information about the measurement settings. It is constructed by a set of operators $\{\meas_k\}_{k=0}^M$ for $M$ different measurement settings. The operators $\meas_k$ depend on the chosen measurement scheme; we will demonstrate how the method works for both homodyne (Sec.~\ref{sec:homodyne}) and heterodyne (Sec.~\ref{sec:heterodyne}) measurements. The elements of the matrix $A$ also depend on a set of $N^2$ basis operators $\{\Omega_k\}_{k=0}^{N^2-1}$ here placed in a one-dimensional array. We use the Fock basis with operators
\begin{equation}
    \Omega_{i\times N + j} = \ketbra{i}{j}, \quad i,j=0, \dots, N-1.
\end{equation}
The density matrix is then represented as $\rho = \sum_{i,j=0}^{N^2-1} \rho_{ij} \Omega_{i\times N + j}$ in this basis.

The matrix $A$ is given by the $M\times N^2$ matrix~%\cite{Teo2021Jun}
\begin{equation}\label{eq:A}
A =
    \begin{pmatrix}
    \Tr[\Pi_0 \Omega_0] & \Tr[\Pi_0\Omega_1] &\hdots &\Tr[\Pi_0\Omega_{N^2-1}] \\
    \Tr[\Pi_1\Omega_0] & \Tr[\Pi_1\Omega_1] &\hdots &\Tr[\Pi_1\Omega_{N^2-1}]\\
    \vdots & \vdots & \ddots & \vdots \\
    \Tr[\Pi_M\Omega_0] & \Tr[\Pi_M\Omega_1] &\hdots & \Tr[\Pi_M\Omega_{N^2-1}]
    \end{pmatrix}.
\end{equation}
%
%overdetermined ~\cite{Sugiyama2013Oct}

%&&&&&&&&&&&&&&&&&&&&&&&&&&&&&&&&&&&&&&&&&&&&&&&&&&&&&&&&&&&&&&&&&&&&&&&&&66
\subsection{The tomography problem}

Direct linear inversion of Eq.~\eqref{eq:linear_eq} to obtain the density matrix elements $\rho_k$ is not feasible since it does not guarantee positive semidefiniteness of the state, so any noise in the measurement data $\vecp$ can result in an unphysical density matrix~\cite{Mogilevtsev1997Nov}. Also, in this case it is likely that no solution exists at all if the system is overdetermined ($M>N$). For this reason it makes sense to determine an approximate solution that renders the residual $\| A\vecrho-\vecp\|$ as small as possible. This means the problem of finding a solution to the set of linear equations $A\vecrho=\vecp$ can be formulated as the optimization problem
%The linear tomography problem can be reformulated as a least-squares problem
%
\begin{equation}\label{eq:leastsquares}
    \min_{\vecrho} \| A\vecrho - \vecp\|^2.
\end{equation}
%  \url{www.cambridge.org/9781107050877}
When the vector norm $\|\cdot\|$ is chosen to be the Euclidean or $\ell_2$-norm~\footnote{For a vector $x \in C^{m}$, the $\ell_{2}$ norm is defined as $\|x\|_{\ell_{2}}=$ $\sqrt{x^{\dagger} x}=\sqrt{\sum_{i=1}^{m}\left|x_{i}\right|^{2}}$.}, this is well-known as the \emph{least-squares problem}~\cite{Boyd2018-ad}. Unlike the linear problem, an analytical solution can be obtained for this problem. However, the ill-posedness of an inverse problem in the form of a linear system of equations~\eqref{eq:linear_eq} is reflected in an \emph{ill-conditioned} matrix $A$. This means an analytical solution by the normal equations is numerically unstable due to the high condition number of $A$, and the result is highly sensitive to noise in the measurement results~\cite{Kress1998}. A stable solution to an ill-posed problem can be obtained via numerical regularization methods~\cite{Engl1996-wx}. Typically this entails adding a term to the expression~\eqref{eq:leastsquares} that is optimized. In our case this would still not be adequate since an unconstrained optimization does not guarantee a physical state.
Fortunately, the quantum tomography problem has a number of nice features that we can utilize as described below.

%%%%%%%%%%%%%%%%%%%%%%%%%%%%%%%%%%%%%%%%%%%%%%%
\subsection{Convex optimization}
% states form a convex set~\cite{Avron2019Jul}

The set of all quantum states is the set of Hermitian matrices with non-negative eigenvalues and unit trace. This is a convex set~\cite{Bengtsson2006May}. The combination of being Hermitian and having non-negative eigenvalues means that a density matrix describing a physical quantum state is positive semidefinite ($\rho \succeq 0$).
As such, we can write our optimization problem as a \emph{semidefinite program}
\begin{align}
        &\text{minimize }\| A\vecrho - \vecp\|^2\text{ over }\rho, \label{eq:cost} \\
        &\text{subject to }\rho \succeq 0, \label{eq:semidef_constraint}\\
        & \hspace{12mm}\Tr\rho =1. \label{eq:trace_constraint}
\end{align}
%
%Solving the problem of minimizing the convex function
The constraints that ensure a physical density matrix also act as regularization, meaning that the problem is now well-posed~\cite{Vogel2006Aug, Souopgui2016Oct, Kirsch2011-zx}.

Since the feasible solution set $\{\rho\}$ is convex, the cost function~\eqref{eq:cost} is a convex function over this set, and the trace equality constraint is linear, it is a convex optimization problem. This means we can utilize efficient convex optimization methods.
%have mosek version 9.2.42
Convex optimization problems are easier to solve than general nonlinear optimization problems, but most importantly, it can be shown that every local minimum to a convex optimization problem is also a global minimum, guaranteeing an optimal solution~\cite{Boyd2004Mar}.

% along with the solver \textsc{mosek}~\cite{mosek}.

%%%%%%%%%%%%%%%%%%%%%%%%%%%%%%%%%%%%%%%%%%%%%%%%%%%%%%%%%%%%%%%%%%%%%%%%%%%%%%%%%%%%%%%%%%%%%%%%%%%%%%%%%%%%%%%%%%%%%%%%%%%%
%\subsubsection{Solution to the convex optimization problem}

The least-squares problem~\eqref{eq:cost} will be the cost function for our convex optimization. The optimization variable corresponding to the unknown state is defined as an $N \times N$ Hermitian matrix $\statevariable$; it is only vectorized when calculating the cost, as indicated by the vector notation $\vecrho$ in Eq.~\eqref{eq:cost} as opposed to $\rho$ in Eqs.~\eqref{eq:semidef_constraint} and~\eqref{eq:trace_constraint}. Having the variable in matrix form makes stating the constraints
\begin{equation}
 \statevariable \succeq 0, \quad \Tr\statevariable = 1
\end{equation}
very straightforward. There are efficient numerical methods that solve these types of problems, and software packages that provide them. We will use the open source Python package \textsc{cvxpy}~\cite{Diamond2016,Agrawal2018Jan, BibEntry2021Jun}. Being able to utilize already existing software makes this state reconstruction procedure very easy to implement.

%%%%%%%%%%%%%%%%%%%%%%%%%%%%%%%%%%%%%%%%%%%%%%%
%\subsection{Convex opt previously}

% almost CV https://link.springer.com/article/10.1007%2Fs10773-006-9287-9
%
% state and meas https://journals.aps.org/pra/abstract/10.1103/PhysRevA.98.042318

%

%%%%%%%%%%%%%%%%%%%%%%%%%%%%%%%%%%%%%%%%%%%%%%%%%%%%%%%%%%%%%%%%%%

\section{Application with common measurement schemes}\label{sec:application}
% detection schemes
%Homodyne detection is a relatively simple and efficient experimental technique for investigating the quantum properties of light %leonarhd. heterodyne = eight-port homodybe detection

Homodyne and heterodyne detection are practical experimental techniques for investigating the quantum properties of light. Below we test the convex optimization state reconstruction method on data from these two measurement schemes, starting with simulated homodyne measurements having different levels of efficiency in section~\ref{sec:homodyne}.
In section~\ref{sec:heterodyne} we move on to heterodyne measurement. First we test the method on simulated noisy data in subsection~\ref{subsec:simulated_noisy_heterodyne}. Then we compare the performance of our method to a neural network and a maximum likelihood method in subsection~\ref{subsec:comp_heterodyne_mle_cgan}, using simulated, ideal data as the available code for those two methods cannot handle noisy data. Finally, we use our method on experimental heterodyne data and compare it to a maximum likelihood method based on moments in subsection~\ref{subsec:exp_heterodyne}. Another example of a reconstruction using experimental data in the form of Wigner tomography can be found in Appendix~\ref{app:wigner}.

%\textcolor{blue}{Here, we show that our simple convex optimization method reconstructs the quantum state from simulated homodyne measurement data faster than the ordinary iterative maximum likelihood method~\cite{Lvovsky2004May} which was implemented in~\cite{Strandberg2019Dec}.}

%In Sec.~\ref{sec:homodyne} we will use simulated homodyne measurement data generated by solving a stochastic master equation, using the same code used in Ref.~\cite{Strandberg2019Dec}, and compare the CXV reconstruction to the iterative maximum likelihood algorithm implemented in the same paper.

%%%%%%%%%%%%%%%%%%%%%%%%%%%%%%%%%%%%%%%%%%%%%%%%%%%%%%%%%%%%%%%%%%
\subsection{Homodyne measurement}~\label{sec:homodyne}

In an optical homodyne experiment, a current proportional to the generalized field quadrature $x_\theta = (a^\dagger \expo{\imi\theta} + a\expo{-\imi\theta})/\sqrt{2}$ is measured, where $a$ and $a^\dagger$ are the bosonic field annihilation and creation operators, respectively. The rotation angle $\theta$ is set by the phase of a local oscillator. The full manifold of quadratures can be obtained by varying $\theta$ between 0 and $\pi$~\cite{Leonhardt1995Jan}. The underlying quantum state of light can be reconstructed from a number of such measurements~\cite{Sych2012Nov}. Here we test the convex reconstruction method with simulated homodyne measurement data generated by solving a stochastic master equation as implemented in Ref.~\cite{Strandberg2019Dec}. Simulated homodyne currents are integrated and the data is discretized by sorting it into bins, here denoted by an index $j$.
Fock basis matrix elements for the measurement operator corresponding to rotation angle $\theta$ and bin $j$ are
\begin{equation}\label{eq:homodyne_pi}
  \Pi_{mn}(\theta,j) = \braket{m|\Pi(\theta,j)|n}=\int_{x_j}^{x_{j+1}}\braket{m|x_\theta}\braket{x_\theta|n}\dif x,
\end{equation}
where the integral from $x_j$ to $x_{j+1}$ is over photocurrents in bin $j$, and the overlap between the number and quadrature eigenstates is given by the harmonic oscillator eigenfunctions in the position basis multiplied by a phase~\cite{barnett2002methods}:
\begin{equation}
  \braket{x_\theta|n}= \expo{-\imi n \theta}\frac{1}{\sqrt{2^n n!}}\left(\frac{1}{\pi}\right)^{1/4} \expo{-x^2/2} H_n(x),
\end{equation}
with $H_n$ being the $n$th Hermite polynomial.

%We use the stochastic master equation and maximum likelihood method implemented in~\cite{Strandberg2019Dec} [Code ref].
We use a test state
\begin{equation}\label{eq:homodyne_teststate}
   \ket{\psi}_\mathrm{homodyne}= \frac{\ket{0} + \ket{2}}{\sqrt{2}},
\end{equation}
because it is a state that is asymmetric in phase space and has Wigner negativity, and it could also be prepared with only small modifications of the code in Ref.~\cite{Strandberg2019Dec}.

%For ideal, noiseless data, we could compare the convex reconstruction to the maximum likelihood method as also implemented in Ref.~\cite{Strandberg2019Dec}. CVX compares favorably, often having a faster running time, but it varies depending on parameters such as Fock space dimension and number of rotation angles. The resulting reconstructions are very close in fidelity.

\subsubsection{State reconstruction with noisy homodyne data}
\begin{figure}[h!]
    \centering
    % 2000 trajs,
    % homodyne/plot_noisy_homodyne_fig.ipynb
    \includegraphics[width=0.65\columnwidth]{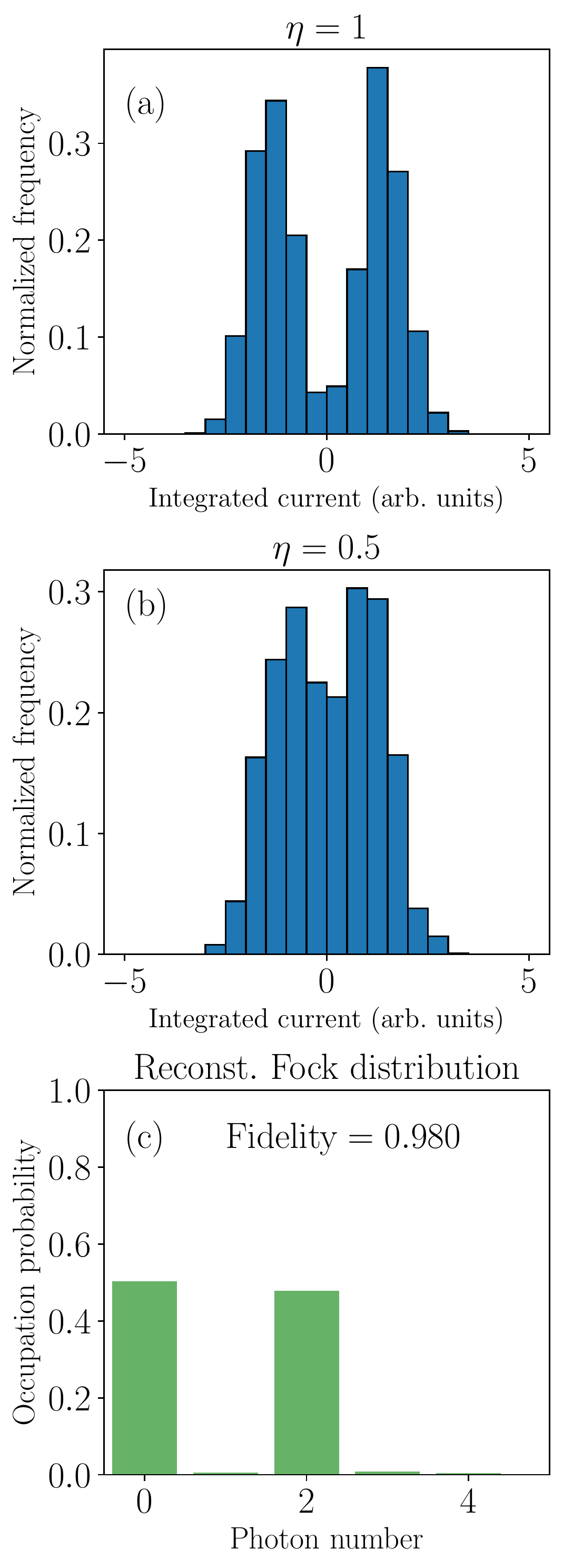}
    \caption{Reconstruction of the state $\ket{\psi}_\mathrm{homodyne}$  [cf.\ Eq.~\eqref{eq:homodyne_teststate}] with low homodyne measurement efficiency $\eta = 0.5$.
    (a) Simulated homodyne current for one of the tomography angles with measurement efficiency $\eta=1$, for comparison. (b) Simulated homodyne current with efficiency $\eta = 0.5$. (c) Photon number distribution of the reconstructed state from one measurement round of noisy $\eta = 0.5$ data.}
    \label{fig:noisy_homodyne_reconst}
\end{figure}

Noise and losses are an inevitable part of any experiment. Common sources are amplification noise and non-ideal detection efficiency, and these types of imperfections can be related to each other as shown in Appendix~\ref{app:equivalence}. In traditional quantum optics literature~\cite{Leonhardt1997-tn} the dominant imperfection is the measurement efficiency $\eta$ of homodyne detection, which is in realistic cases typically less than 1, where $\eta=1$ would correspond to \SI{100}{\%} efficiency. Inefficiency in homodyne tomography can be accounted for in the reconstruction by modifying the measurement operator as~\cite{Lvovsky2004May}
\begin{equation}
\begin{split}
        \tilde\Pi &= \sum_{m,n,k} B_{m+k,m}B_{n+k,n} \Pi_{mn}(\theta,j) \ketbra{n+k}{m+k}.
\end{split}
\end{equation}
with
\begin{equation}
    B_{n+k,n}(\eta) = \sqrt{\binom{n+k}{n}\eta^n(1-\eta)^k.
    }
\end{equation}
\begin{table}[h!]
% homodyne/test_efficiency/noisy_homodyne.py
\centering
\begin{tabular}{ c | c }
 $\eta$ & Fidelity\\
 \hline
 1.0 & 0.995 \\
 0.9 &  0.990  \\
 0.8 &  0.985\\
 0.7 &  0.991  \\
 0.6 &  0.976 \\
 0.5 &  0.985  \\
 0.4 &  0.940  \\
 0.3 &  0.913 \\
  0.2 & 0.758\\
 0.1 & 0.711
\end{tabular}
\caption{Fidelities between the reconstructed state and the test state~\eqref{eq:homodyne_teststate} for different levels of measurement efficiency $\eta$, where $\eta=1$ is perfect efficiency.
}\label{table:fidelities}
\end{table}
Using this, our reconstruction of the test state~\eqref{eq:homodyne_teststate} works well even with very inefficient measurements, with fidelities over 0.9 for efficiencies as low as $\eta=0.3$ (see Table~\ref{table:fidelities}), where the fidelity between the reconstructed state $\rho$ and the ideal state $\rhoideal$ is defined as
\begin{equation}
    %\mathrm{Fidelity}
    F(\rho, \rhoideal) =\left( \Tr \sqrt{\sqrt{\rho}\rhoideal\sqrt{\rho}}\right)^2.
\end{equation}
The simulated measurement used 20 rotation angles, with data binned into 20 bins per angle. Due to fluctuations in the simulated data which consists of integrated currents for 2000 stochastic trajectories~\cite{Wiseman1993Jan}, the values are averaged over six measurement rounds.
A reconstruction of the test state from data with efficiency $\eta=0.5$ is displayed in Fig.~\ref{fig:noisy_homodyne_reconst}.

To the best of our knowledge, there is unfortunately no publicly available code for other reconstruction methods using noisy homodyne data to compare with.
The code and simulated data used to produce these results are provided in Ref.~\cite{code}.

%&&&&&&&&&&&&&&&&&&&&&&&&&&&&&&&&&&&&&&&&&&&&&&&&&&&&&&&
\subsection{Heterodyne measurement}\label{sec:heterodyne}

Noiseless heterodyne measurement corresponds to measuring the Husimi $Q$-function. The $Q$-function for a state $\rho$ is defined as~\cite{Rundle2021Jun}
\begin{equation}
    Q(\alpha) = \frac{1}{\pi} \braket{\alpha|\rho|\alpha} = \frac{1}{\pi} \Tr[\ketbra{\alpha}{\alpha}\rho].
\end{equation}
The corresponding measurement operators are projections onto coherent states $\Pi_k=\ketbra{\alpha_k}{\alpha_k}$ where $\alpha_k \in\mathbb{C}$ denotes a point in phase space.
We test the method on a coherent superposition of coherent states $\ket\beta$, also called Schrödinger's cat state:
\begin{equation}\label{eq:teststate_cat}
    \ket{\psi}_\mathrm{heterodyne}=\frac{\ket{\beta}+\ket{-\beta}}{\sqrt{2}}, \quad \beta=2.
\end{equation}
It is an interesting looking state [cf.\ Fig.~\figpanel{fig:noisy_cat}{c}] of interest for error-corrected quantum computing~\cite{Mirrahimi2014Apr}. The code used to produce the results below are provided in Ref.~\cite{code}.

%-----------------------------------
\subsubsection{State reconstruction with noisy heterodyne data}\label{subsec:simulated_noisy_heterodyne}

As an example of a noisy system, consider superconducting circuits, which provide a promising platform for quantum computing experiments. Since those circuits must be cooled to extremely low temperatures, thermal noise is a disruptive factor. It can be generated at room temperature and transmitted to the chip, but additionally, amplification of a weak measurement signal inevitably adds noise. This noise can be modeled as a thermal state~\cite{Clerk2010Apr}.
When a measurement contains amplifier noise described by a thermal state \rhotherm{n} with mean number of photons $n$, the measured histogram corresponds to the convolution~\cite{Kim1997Oct}
\begin{equation}\label{eq:noise_conv}
    \int \ptherm{n}(\alpha^* - \beta^*) Q(\beta)\, \dif{}^2\beta,
\end{equation}
where $Q(\beta)$ corresponds to the Husimi $Q$-function~\cite{Cahill1969Jan} of the underlying wanted state, and 
\begin{equation}\label{eq:p_func_thermal}
    \ptherm{n} (\alpha)= \frac{1}{\pi n}\expo{-|\alpha|^2/n},
\end{equation}
is the $P$-function of the thermal state~\cite{Glauber1963Sep}. The noise-compensated measurement operators are given by~\cite{Eichler2013}
\begin{equation}~\label{eq:noisecomp_pi_homodyne}
    {\tilde \Pi^k}_n  = D(\alpha_k) \rhotherm{n} D^\dagger(\alpha_k).
\end{equation}

\begin{figure}[h!]
    \centering
    \includegraphics[width=0.8\columnwidth]{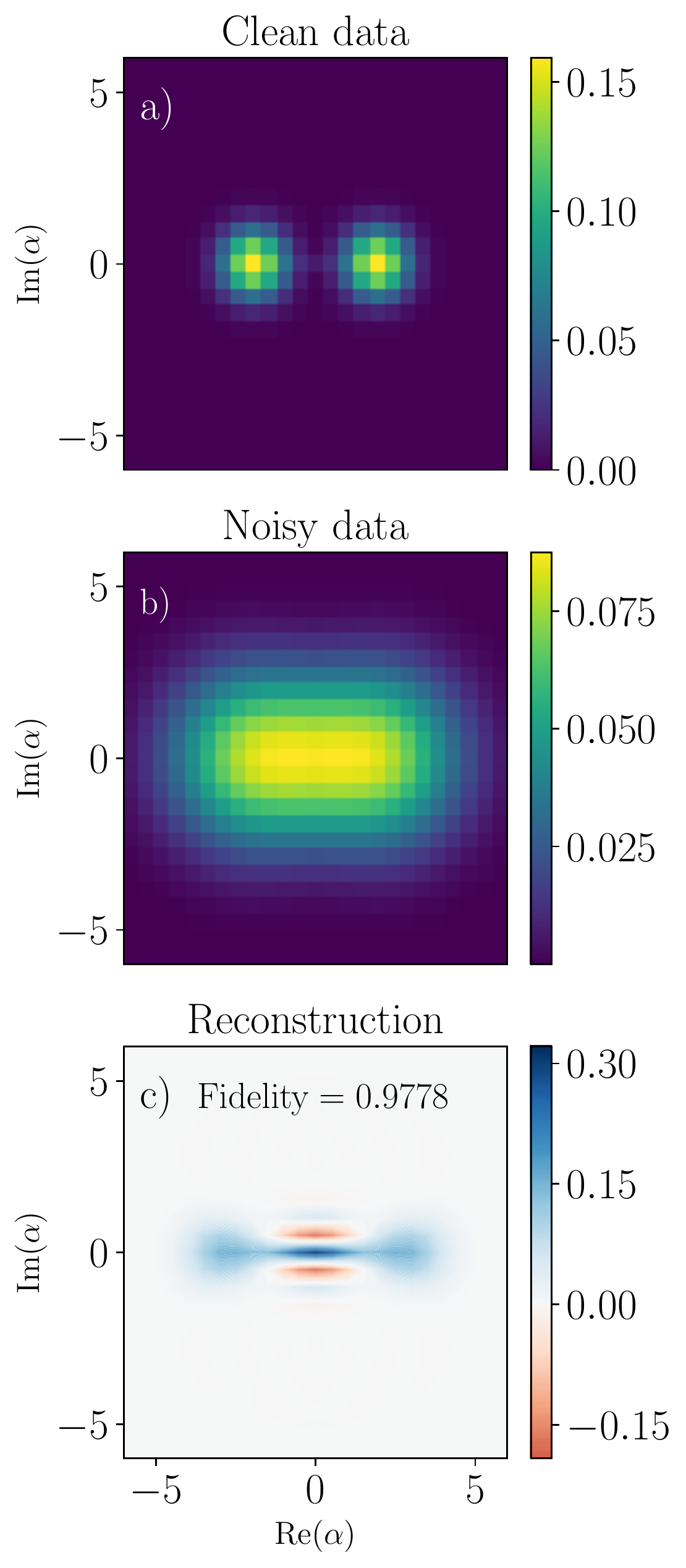}
    \caption{Reconstruction of test state $\ket{\psi}_\mathrm{heterodyne}$ [cf.\ Eq.~\eqref{eq:teststate_cat}] from noisy data. Parameters are $N=32$, $25\times 25$ grid wih phase space limits $\xmax=\pmax=6$. a) Ideal noiseless data corresponding to the Q-function. b) Data corrupted by thermal noise corresponding to $\nthermal=5$. c) Wigner function of the reconstructed state. The reconstruction fidelity is very high despite the noisy data.
}
    \label{fig:noisy_cat}
\end{figure}

For testing, simulated noiseless data was obtained by calculating the $Q$-function of the test state at discrete points on a grid in phase space. This was then convolved with the thermal $P$-function to generate simulated noisy histograms.
%
%Superconducting circuits is a promising platform for quantum computing experiments. Since those circuits must be cooled to extremely low temperatures, thermal noise is a disruptive factor. It can be generated at room temperature and transmitted to the chip, but additionally, amplification of a weak measurement signal inevitably adds noise. This noise can be modeled as a thermal state~\cite{Clerk2010Apr}.
%
Using the noise-compensated operators~\eqref{eq:noisecomp_pi_homodyne} to construct the matrix $A$ defined in Eq.~\eqref{eq:A}, we show in Fig.~\ref{fig:noisy_cat} that the underlying quantum state can be reconstructed with very high fidelity from noisy data. The clean histogram is displayed in Fig.~\figpanel{fig:noisy_cat}{a} for comparison to the noisy histogram Fig.~\figpanel{fig:noisy_cat}{b} which is smeared out due to thermal noise with $\nthermal=5$ photons on average. The state was reconstructed using the noisy data and the knowledge that $\nthermal=5$. The high-fidelity result can be seen in Fig.~\figpanel{fig:noisy_cat}{c}. This is a reconstruction the neural network in Ref.~\cite{Ahmed2021Sep} could not do.
%
% noisy_cat_plot.ipynb
%
%This means elements of $b$ can be measured by heterodyne detection.

%-------------------------------------------------------------------------------
\subsubsection{Comparison to maximum likelihood estimation and neural network}\label{subsec:comp_heterodyne_mle_cgan}

\begin{figure}[h!]
    \centering
    % /cvx_state_tomography/comparison_notebooks/qst-cvx/notebooks/read_test_all.py
    \includegraphics[width=0.9\columnwidth]{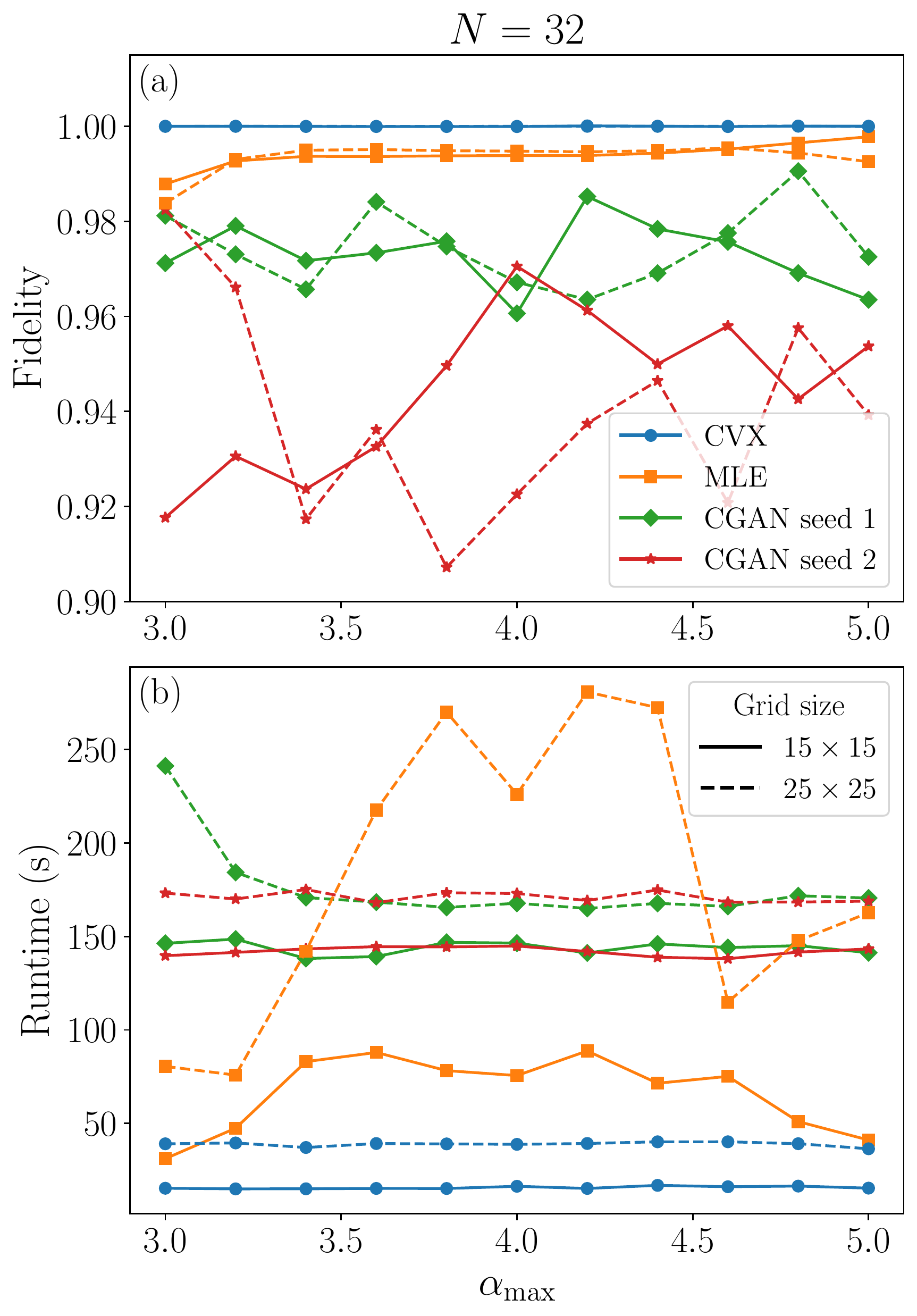}
    \caption{Comparison between reconstruction methods using simulated ideal data. The plots show fidelities and runtimes for reconstruction of the test state~\eqref{eq:teststate_cat} for different phase space limits $\amax$, and grid sizes, with a fixed Fock space dimension $N=32$. The method is indicated by color and markerstyle, while the grid size is indicated by linestyle (solid for $15\times 15$, dashed for $25\times25$) in both subplots. (a) Fidelities. CVX is very stable at fidelity 1 for all parameter combinations and both grid sizes. MLE is fairly stable around 0.99. The CGAN fidelity behaves erratically. It tends to need a finer grid to reach high fidelity. (b) Runtimes. CVX is almost always fastest. The convergence time of MLE tends to vary, while the CGAN reconstruction time is fairly constant.}
    \label{fig:different_amax}
\end{figure}

We compare our convex optimization method (CVX) to a standard maximum likelihood estimation method (MLE) and a recent method based on a type of neural network called conditional generative adversarial network (CGAN)~\cite{Ahmed2021Sep,Ahmed2021Sep1} with code provided in Ref.~\cite{shahnawaz_ahmed_2021_5105470}.
We observe that MLE and the CGAN can behave erratically depending on reconstruction parameters for the same underlying state, even without noise. For example, we show a test with different phase space limits $\amax=\xmax=\pmax$ and two different grid sizes in Fig.~\ref{fig:different_amax}.
%The state is the same, but the density of data is modified with different $\amax$ and grid sizes. For a fixed grid size, larger $\amax$ means measurements are more sparse in phase space. Correspondingly, for a fixed $\amax$, a larger grid means higher density of measurements.
%
%
%In the source code the reconstruction terminates after 1000 iterations.
The MLE is set to terminate when the Hilbert-Schmidt distance between two consecutive estimations is smaller than $10^{-6}$. The fidelity is fairly stable, but as seen in Fig.~\figpanel{fig:different_amax}{b}, it is generally the slowest method with the larger grid. The CGAN is initialized with random parameters for each training, leading to different results for each run. In Fig.~\figpanel{fig:different_amax}{a} we show the result for two different random seeds. Not only does the fidelity vary for different seeds, it fluctuates significantly for different $\amax$.
Notably, our CVX method consistently reconstructs the state with fidelity 1, almost always with the fastest runtime. For more information on the time complexity of the CVX method, see Appendix~\ref{app:comp_cost}.% As can be expected, reconstruction with the larger grid takes a longer time in general.

\begin{figure}[h!]
    \centering
    %~/cvx_state_tomography/comparison_notebooks/qst-cvx/notebooks$ python read_test_N.py
    \includegraphics[width=0.85\columnwidth]{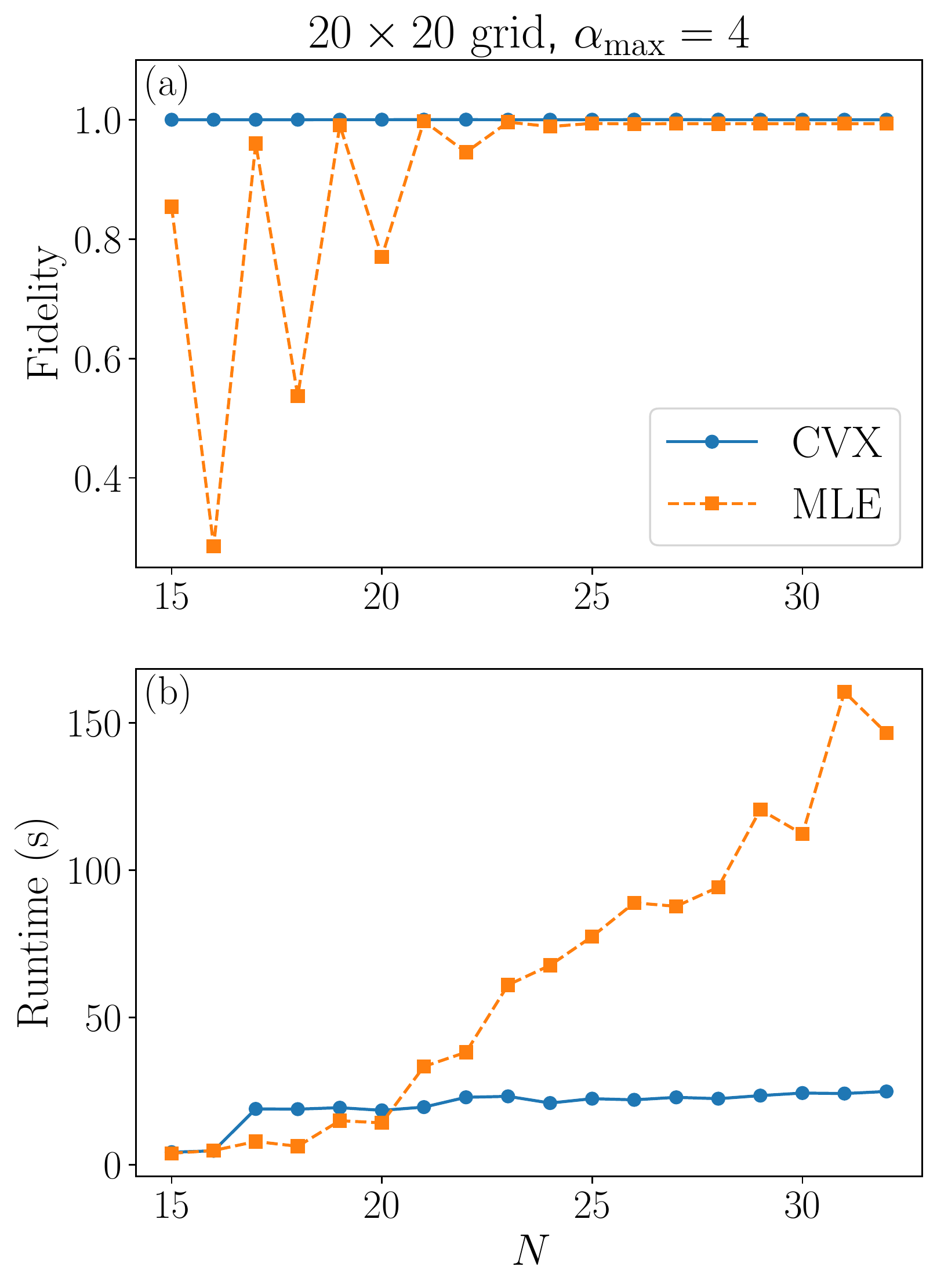}
    \caption{Comparison between reconstruction methods using simulated ideal data. The plots show fidelities and runtimes for reconstruction of the test state~\eqref{eq:teststate_cat} for different Fock space dimensions $N$, with a fixed $20\times 20$ grid and phase space limit $\amax=4$. (a) Fidelities. CVX (circle marker and solid line) is stable at fidelity 1 for all values of $N$ while the MLE (square marker and dashed line) fidelity starts to jump unexpectedly for $N<23$. (b) Runtimes. For $N>20$ CVX is faster than MLE.}
    \label{fig:N_cvx_mle}
\end{figure}

We also test reconstruction using different Fock space dimensions $N$. Since the CGAN is fixed at $N=32$ we only compare CVX and MLE in Fig.~\ref{fig:N_cvx_mle}. The CVX reconstruction is stable at fidelity 1, but the MLE reconstruction starts to deviate in an unexpected manner for $N<23$, even though the test state is fully contained in Fock spaces down to dimension 15. The figure shows reconstructions with a fixed $20\times 20$ grid, but the same behavior was observed for other grid sizes. These comparisons were performed with the ideal $Q$-function. To see how sampling errors can affect the fidelies of CVX and MLE reconstructions, go to Appendix~\ref{app:stats}.

%%%%%%%%%%%%%%%%%%%%%%%%%%%%%%%%%%%%%%%%%%%%%%%%%%%%%%%%%%%%%%
\subsubsection{State reconstruction with experimental heterodyne data}\label{subsec:exp_heterodyne}

% all 15 N=8 recons took 6 and a half minutes.

\begin{figure}[h!]
    \centering
    % plot_cvx_mle_fidelities.py from data_heterodyne_testnoise_getstate.py
    %cvx_state_tomography/data_yong_W/500TracesPerHist20MHz_Gain=121.73/plot_cvx_mle_fidelities.py
    % edited in inkscape
    \includegraphics[width=\columnwidth]{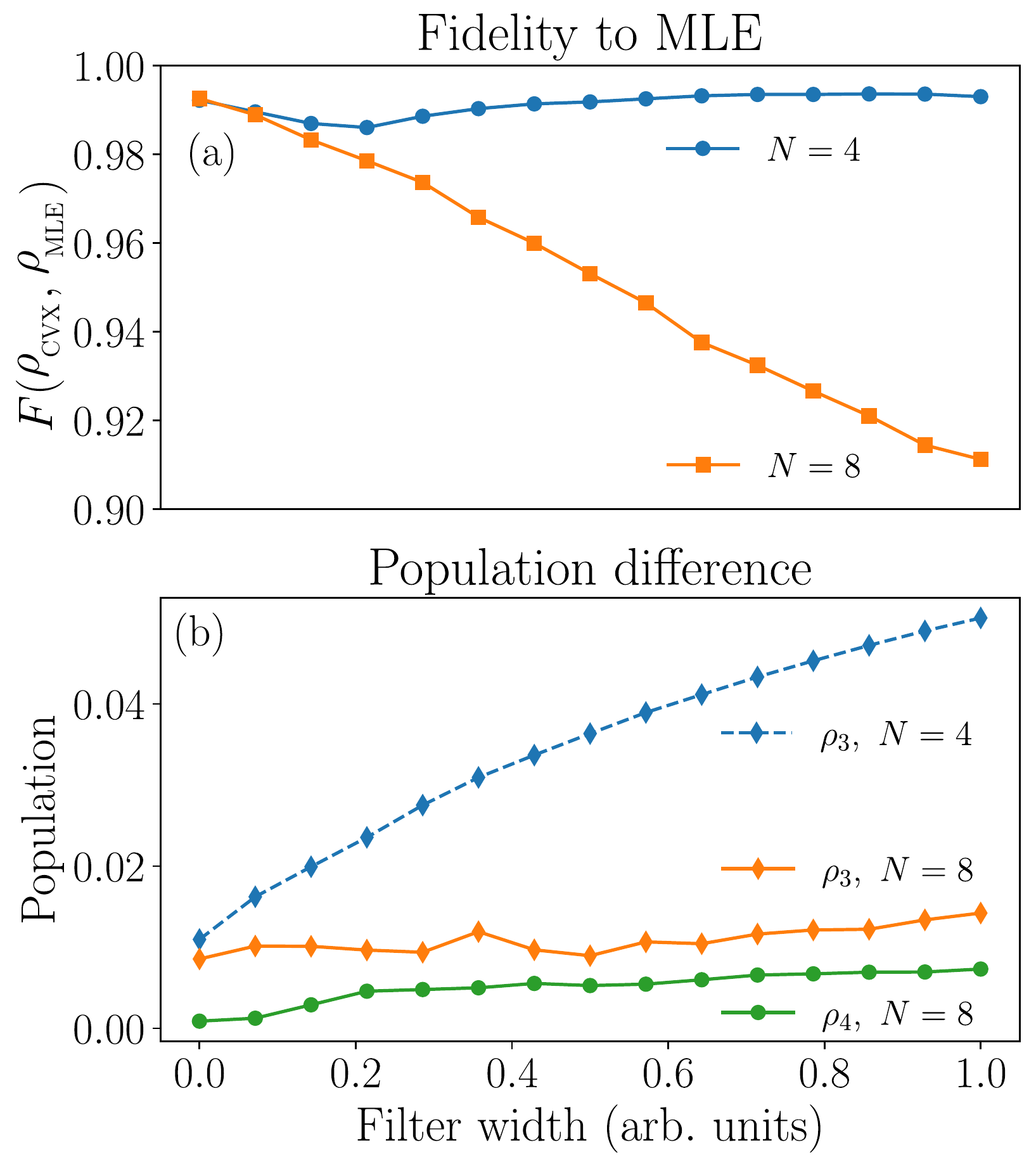}
    \caption{Comparison between moments-MLE reconstructions with $N=4$ and CVX reconstructions with $N=4$ and $N=8$ for experimental data with different filter widths, corresponding to states with different photon numbers. (a) Fidelity between reconstructions. When CVX is truncated to the same Fock space dimension ($N=4$) as the MLE, fidelity is close to 1, meaning the two reconstruction methods agree. When CVX is allowed a larger Fock space, fidelity decreases with longer temporal filters, corresponding to states with higher photon numbers. This indicates that $N=4$ is not a large enough Fock space to reconstruct the correct state. Note that the $y$-axis begins at 0.9.\ (b) Three-photon populations $\rho_3$ for the $N=4$ and $N=8$ reconstructions, and four-photon population $\rho_4$ for the latter. It can be seen that for $N=4$, the largest possible photon number state $\rho_3$ increases excessively when higher photon number populations instead start appearing with the larger Fock space.}
    \label{fig:yong_exp}
\end{figure}

Here we demonstrate the CVX state reconstruction method on experimental data produced for the publication Ref.~\cite{Lu2021Jun}. The experiment consisted of selecting a particular mode from a propagating field which was emitted by a coherently driven qubit in front of a mirror. Particular modes were chosen by the shape of a temporal filter. The procedure for state reconstruction with noisy data used in this paper, and commonly used in circuit QED, was to extract moments of the bosonic operators $\hat a$, $\hat a^\dagger$ from the heterodyne histogram and then doing a maximum likelihood estimation of the state based on the extracted moments~\cite{Eichler2011Jun, Eichler2012Sep, Lu2021Sep}. For states with large photon numbers, high-order moments are required, since $n$th order moments only contain information about number states up to $n-1$. This means full information of a density matrix in a Fock space of dimension $N=n-1$ can be obtained. This method is however limited to low photon-number states, since the standard deviations of the moments increase rapidly with higher and higher moments. For this reason, the Fock space was truncated to $N=4$ in Ref.~\cite{Lu2021Jun}.
Notably, this is not a restriction for the CVX method. In Fig.~\figpanel{fig:yong_exp}{a} we show the fidelities between the moments-based MLE with $N=4$ and CVX with $N=4$ and $N=8$, where the fidelity for the latter was calculated by truncating the reconstructed density matrix. The $x$-axis in Fig.~\ref{fig:yong_exp} corresponds to the filter width, where larger widths are expected to correspond to states with higher photon numbers. Each filter width corresponds to separate states with separate measurement histograms. In Fig.~\figpanel{fig:yong_exp}{a} it can be seen that the fidelity between the MLE and CVX reconstructions with both $N=4$ is close to 1 for all measurements, indicating that CVX performs at least as well as MLE. However, if the CVX reconstruction Fock space is increased to $N=8$, fidelity starts to decrease for longer filter widths, i.e.\ states with higher photon numbers. In Fig.~\figpanel{fig:yong_exp}{b} we see that for the more severely truncated $N=4$ density matrix, the three-photon population $\rho_3$ increases in lieu of higher photon-number populations. But with $N=8$, we see that the four-photon probability $\rho_4$ (and to an even lesser degree higher photon-number states not shown in the plot) is slightly increasing. Because the CVX reconstruction with $N=8$ matches the reconstruction with $N=4$ with high fidelity for the shortest temporal filters, it is reasonable to suspect that the smaller density matrix from the moments-MLE method fails to completely correctly capture the states with a slightly larger photon number while the CVX method has no such issue.% Running all reconstruction for 15 filter widths with $N=8$ only took six an a half minutes on a computer with 16 cores.

%%%%%%%%%%%%%%%%%%%%%%%%%%%%%%%%%%%%%%%%%%%%%%%%%%%%%%%%%%%%%%
\section{Summary and conclusions}\label{sec:conclusions}

Convex optimization for quantum tomography is not a new idea in itself, but it does not seem to have been implemented for single-mode continuous variable state tomography previously.Unlike emerging machine learning methods, the convex optimization method is easy to understand and to use. Here we have demonstrated such a method on different types of realistic, noisy data. We showed that the convex optimization method gives stable results even when measurement parameters are varied, unlike an implementation of maximum likelihood estimation as well as a neural network, which both gave aberrant reconstructions for certain parameter combinations seemingly without reason. Besides the reliability of the convex optimization reconstruction, a major benefit is that unlike iterative maximum likelihood and neural network methods, there is no arbitrary stopping criteria for when to terminate the iterations, and the optimal solution is guaranteed to be reached. The method is easy to implement with openly available Python packages, and example code is provided in Ref.~\cite{code}.

While continuous-variable states was the focus of this study, the presented framework and method is general and can also be used for discrete-variable systems. The state reconstruction takes less than 30 seconds for states occupying a Hilbert space of dimension $N$ up to 32 [cf. Fig.~\figpanel{fig:N_cvx_mle}{b}], and under ten minutes with $N=64$ on a desktop computer with 16 cores. Based on this, the current method is expected to be practical for states with up to six qubits. It is notable that the method is not limited by the computational cost of the convex optimization itself, but rather by the construction of the necessary operators (see Appendix~\ref{app:comp_cost}). This was not the main concern of this work, so there is possibly room for improvement.

%Heterodyne measurements correspond to measuring the Husimi $Q$-function $Q(\alpha)=\frac{1}{\pi}\braket{\alpha|\rho|\alpha}$ \textcolor{blue}{for $\eta$=1}
%%%%%%

%%%%%%%%%%%%%%%%%%%%%%%%%%%%%%%%%%%%%%%%%%%%%%%%%%%%%%%%%%%%%%%%
%%%%%%%%%%%%%%%%%%%%%%%%%%%%%%%%%%%%%%%%%%%%%%%%%%%%%%%%%%%%%%%
\begin{acknowledgments}

Huge thanks to Shahnawaz Ahmed for valuable discussions about all things tomography as well as programming.
Also big thanks to Yong Lu and Marina Kudra for providing experimental data.
Thanks to Christopher Eichler for an informative discussion.
Finally I thank Fernando Quijandr\'{\i}a for discussions and proofreading the manuscript, and Göran Johansson for proofreading as well. This work was supported by Chalmers Excellence Initiative Nano.
%Chalmers Excellence Initiative Nano is acknowledged for financial support.

\end{acknowledgments}

\appendix

%&&&&&&&&&&&&&&&&&&&&&&&&&&&&&&&&&&&&&&&&&&&&&&&&&&&&&&
\section{Vectorization of $\rho$}\label{app:vectorize}

The density matrix is cast to a vector in row-major order, meaning that the second column is placed under the first, and so on. As an example, we show the vectorization of a density matrix with $N=2$:
\begin{equation}
    \rho =
    \begin{pmatrix}
    \rho_{00} & \rho_{01} &  \rho_{02} \\
        \rho_{10} & \rho_{11} &  \rho_{12} \\
    \rho_{20} & \rho_{21} &  \rho_{22} \\
    \end{pmatrix}.
\end{equation}
The vectorized version is
\begin{equation}
\vecrho =
    \begin{pmatrix}
    \rho_{00} \\
    \rho_{10} \\
    \rho_{20} \\
    \rho_{01} \\
    \rho_{11}\\
    \rho_{21}\\
    \rho_{02} \\
    \rho_{12} \\
    \rho_{22}
    \end{pmatrix}.
\end{equation}

%%%%%%%%%%%%%%%%%%%%%%%%%%%%%%%%%%%%%%%%%%%%%%%%%%%%%%%%%%%%%%%%%%
\section{State reconstruction with Wigner tomography}\label{app:wigner}

The convex optimization state reconstruction can also be performed for Wigner tomography. The Wigner function can be measured as the expectation value of the displaced parity operator $P$~\cite{Royer1977Feb,Banaszek1999Jul}:
\begin{equation}\label{eq:wigner}
    W(\alpha) = \frac{2}{\pi} \Tr[D(\alpha) P D^\dagger(\alpha) \rho],
\end{equation}
with the displacement operator $D(\alpha)=\expo{\alpha \hat a^\dagger - \alpha^*\hat a}$, where $\hat a$ and $\hat a^\dagger$ are the annihilation and creation operators of the bosonic mode. From Eq.~\eqref{eq:wigner} the corresponding measurement operators are seen to be $\Pi_k=D(\alpha_k)P D^\dagger(\alpha_k) $.

\begin{figure}[h!]
% example_notebooks/data_cvx_wigner_tomography.ipynb
    \centering
    \includegraphics[width=1\columnwidth]{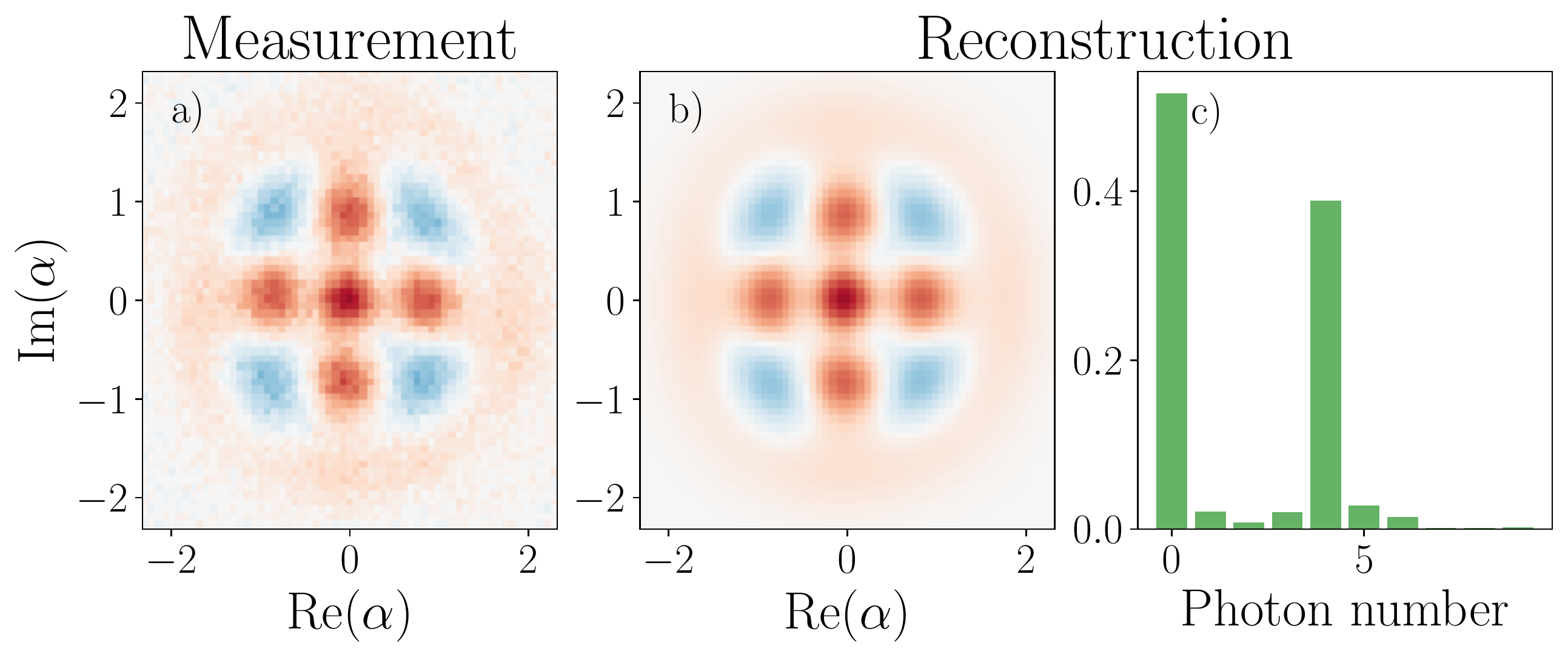}
    \caption{Wigner tomography of the target state $(\ket 0 + \ket 4)/\sqrt{2}$. a) Measurement data. b) Wigner function of the reconstructed state. c) Photon populations of the reconstructed state.  }
    \label{fig:data_marina}
\end{figure}

We test the method on measurement data of the target state $(\ket 0 + \ket 4)/\sqrt{2}$ as shown in Fig.~2(b) in Ref.~\cite{Kudra2021Nov}, recreated in Fig.~\figpanel{fig:data_marina}{a}. The Wigner function and photon populations of the reconstructed state are shown in Figs.~\figpanel{fig:data_marina}{b} and~\figpanel{fig:data_marina}{c}, respectively. The phase space grid is $61\times 61$ with limit $\amax=2.32$, and the reconstruction was performed with Fock space dimension $N=10$. Unit fidelity is not expected due to losses, and our reconstructed state has 0.95 fidelity to the target state, which is close to 0.94 as reported in the paper. The reconstruction in~\cite{Kudra2021Nov}, was produced by the CGAN~\cite{Ahmed2021Sep, Ahmed2021Sep1}-%, and takes around one minute to run. The CVX reconstruction is even faster, taking just over 11 seconds using only a single core (no parallellization).

%%%%%%%%%%%%%%%%%%%%%%%%%%%%%%%%%%%%%%%%%%%%%%%%%%%%%%%%%%%%%%%%%%
\section{Equivalence of thermal noise and detection inefficiency}\label{app:equivalence}

The expression for a heterodyne histogram corrupted by $n$ thermal photons is given by inserting Eq.~\eqref{eq:p_func_thermal} into Eq.~\eqref{eq:noise_conv}, which results in the Gaussian convolution
\begin{equation}\label{eq:p_conv}
    \frac{1}{\pi n}\int \expo{-\dfrac{|\alpha^* - \beta^*|^2}{n}} Q(\beta)\, \dif{}^2\beta.
\end{equation}
We now introduce the $s$-parametrized quasiprobability distribution $\sparam{\alpha}{a}$ of which the $Q$-function, $P$-function and the Wigner function are special cases of for $s=-1, 1, 0$, respectively. For different values of $s$ the distributions are related to each other via a Gaussian convolution~\cite{Cahill1969Jan}
\begin{equation}\label{eq:s_conv}
    \sparam{\alpha}{s} = \frac{1}{\pi}\frac{2}{t-s}\int \expup{-\dfrac{2|\alpha-\beta|^2}{t-s}} \sparam{\beta}{t} \dif{}^2\beta.
\end{equation}
To show the equivalency between thermal noise and detection inefficiency we utilize that there is a relation between the parameter $s$ and the quantum efficiency $\eta$~\cite{Leonhardt1993Dec}:
\begin{equation}
    s = - \frac{2-\eta}{\eta}.
\end{equation}
Inserting this and also setting $t=-1$ in Eq.~\eqref{eq:s_conv} gives
\begin{equation}\label{eq:eta_s_conv}
   \frac{1}{\pi} \frac{2}{\frac{2-\eta}{\eta}-1} \int  \expup{-\dfrac{2|\alpha-\beta|^2}{\frac{2-\eta}{\eta}-1}}  Q(\beta)  \dif{}^2\beta.
\end{equation}
Comparing Eqs.~\eqref{eq:p_conv} and~\eqref{eq:eta_s_conv} we can after some simple algebra see that there is a relation
\begin{equation}
    n = \frac{1}{\eta}-1,
\end{equation}
or equivalently
\begin{equation}
    \eta = \frac{1}{n+1}.
\end{equation}

The equivalence between the models of loss and noise can intuitively be understood by imagining a fictitious beam splitter in front of a perfect detector. The transmissivity $\eta$ of the beam splitter corresponds to the detection efficiency. Any and all types of losses can be modelled in this fashion; the \emph{overall} detection efficiency $\eta$ is simply the product of efficiencies of different parts of the system. Also, dissipation is intimately related to fluctuations~\cite{Kubo1966Jan}, and this statistical noise is formally described by vacuum entering via the second port of the beam splitter. In the case of additional thermal noise, a thermal field can be imagined to enter the second port of the beam splitter~\cite{Leonhardt1997-tn}.

%%%%%%%%%%%%%%%%%%%%%%%%%%%%%%%%%%%%%%%%%%%%%%%%%%%%%%%%%%%%%%%%%%
\section{Notes on the computational cost}\label{app:comp_cost}

CVXPY is a modeling language for convex optimization problems~\cite{BibEntry2021Jun,Diamond2016}, not a solver in itself. CVXPY supports several solvers, and the computational complexity may be different for different solvers. The results in this paper were obtained with the Splitting Conic Solver (SCS)~\cite{ocpb:16,scs}. The SCS algorithm comprises several steps, but the most time-consuming operation is projection on the semidefinite cone which has cubic complexity with respect to the matrix size~\cite{Rontsis2022Jan}. The SCS method is constructed to handle very large problems and is indeed very efficient for the tomography problem. In fact, constructing the matrix $A$~\eqref{eq:A} is by far the most costly part of the CVX tomography program rather than the convex optimization itself. This is exemplified in Table~\ref{table:comp_scaling}, which shows computation times for different Hilbert space dimensions $N$.
\begin{table}[h!]
%~/cvx_state_tomography/comparison_notebooks/qst-cvx/notebook timing_comparison4.dat
\centering
\begin{tabular}{ c | c | c}
 $N$ & $T_\text{ops}$ (s) & $T_\text{sol}$ (s)\\
 \hline
 20 & \SI{18.9}{} & \SI{0.8}{} \\
 25 & \SI{19.0}{} & \SI{1.1}{} \\
30 & \SI{21.4}{} & \SI{1.4}{} \\
35 & \SI{89.2}{} & \SI{1.9}{} \\
40 & \SI{96.7}{} & \SI{2.4}{} \\
45 & \SI{115.7}{} & \SI{2.9}{} \\
50 & \SI{362.5}{} & \SI{3.5}{} \\
55 & \SI{400.8}{} & \SI{4.0}{} \\
60 & \SI{446.4}{} & \SI{4.4}{} \\
\end{tabular}
\caption{Times $T_\text{ops}$ to construct operators and times $T_\text{sol}$ for the convex solver of a CVX tomography program with different Hilbert space dimensions $N$. The parameters are $20 \times 20$ measurement points within $\amax=4$. It can be seen that the convex optimization is very fast, and the overall state reconstruction speed is limited by the operator construction.
}\label{table:comp_scaling}
\end{table}
%
%The part of the program with the largest time-complexity is construction of the matrix $A$~\eqref{eq:A} from the measurement operators $\Pi$ and basis operators $\Omega$.
%
Creating one element of $A$ consists of multiplication between the $N \times N$ matrices $\Pi$, $\Omega$, which scales as $O(N^3)$~\cite{Allaire2008}, and taking the trace which scales linearly. As such, the overall time complexity of this step is $O(N^3)$. This is to be done for each of the $MN^2$ elements of $A$, where $M$ is the number of measurements. As such, the cost of constructing the entire matrix is $O(N^4M)$. The process can be parallelized to improve the speed, but even using 16 cores the matrix construction is the limiting factor of the CVX tomography program as can be seen in Table.~\ref{table:comp_scaling}. It is possible this can be sped-up further, for example by using GPUs.

%%%%%%%%%%%%%%%%%%%%%%%%%%%%%%%%%%%%%%%%%%%%%%%%%%%%%%%%%%%%%%%%%%
\section{Sampling \& statistics: comparison with MLE}\label{app:stats}

\begin{figure}[htb!]
    \centering
    \includegraphics[width=0.8\columnwidth]{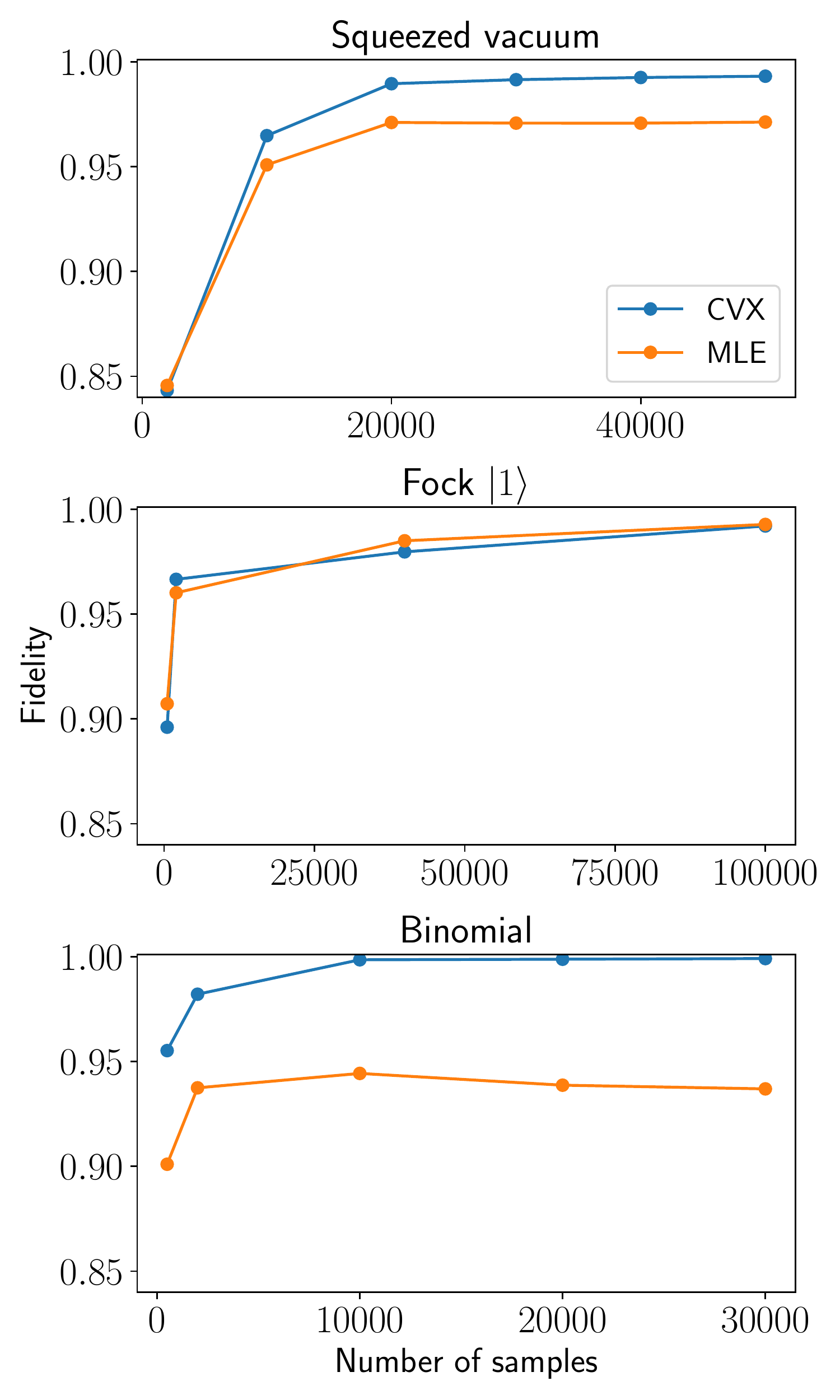}
    \caption{Fidelity of state reconstructions for three states with CVX and MLE, with different numbers of sampled data [cf.\ Fig.~\ref{fig:sampled_histograms}]. Note the different $x$-axis limits for the three plots. }
    \label{fig:compare_sampling}
\end{figure}

\begin{figure}[htb!]
    \centering
    \includegraphics[width=\columnwidth]{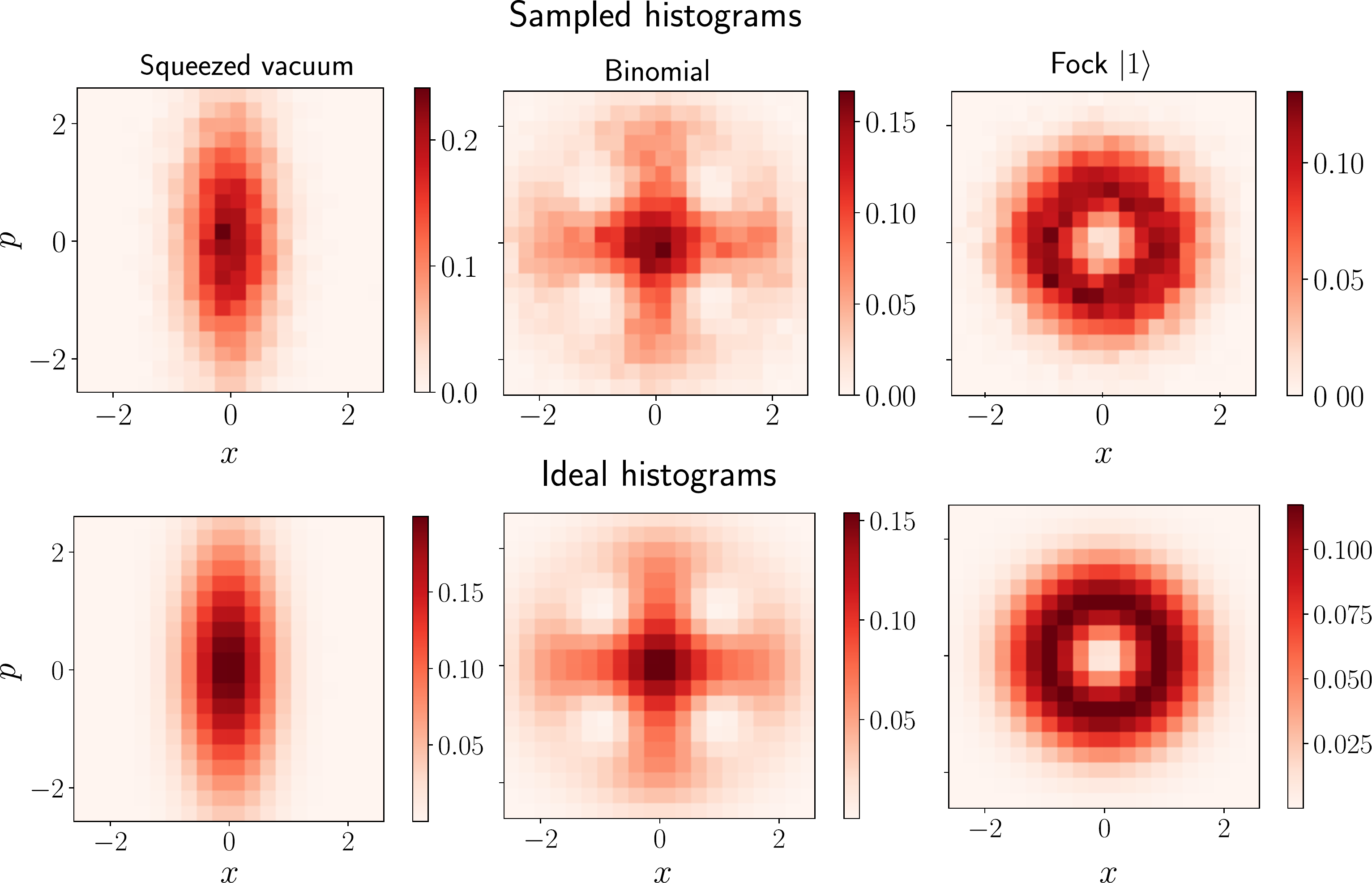}
    \caption{Examples of sampled histograms with 20 000 samples, and the ideal histograms which are the discretized $Q$-functions corresponding to infinite samples.}
    \label{fig:sampled_histograms}
\end{figure}

It is difficult to ascertain the exact number of measurement samples that is needed to obtain sufficient statistics, since this is state dependent as can be seen in Fig.~\ref{fig:compare_sampling}. Nevertheless, a comparison can be made between reconstruction methods. The figure shows reconstruction fidelities as a function of the number of samples for three different states: squeezed vacuum, a so-called binomial code state $(\ket{0} + \ket{4})/\sqrt{2}$ and the Fock state $\ket 1$. The samples are collected into $20 \times 20$ bins, and examples of histograms with 20 000 samples are shown in Fig.~\ref{fig:sampled_histograms}. Example code is provided in Ref~\cite{code} to facilitate further testing.

%projection has cubic complexity wrt matrix size 
%\url{https://link.springer.com/article/10.1007/s10957-021-01971-3}
%Following an initial factorization of an m×m matrix, every iteration of ADMM entails the solution of a linear system via forward/backward substitution and a projection to the Semidefinite Cone. For SDPs, this projection operation typically takes the majority of the solution time, sometimes 90\% or more

%faster ADMM \url{https://link.springer.com/article/10.1007/s10957-021-01971-3}
%normal is cubic

%Operator construction is more costly
%even when parallelized on 16 cores

%Matrix multiplication $O(N^3)$ and then taking the trace which is $O(N)$. This is done for each element in $A$, of which there are $MN^2$. This results in a time complexity $O(MN^4)$
%\url{https://link.springer.com/book/10.1007/978-0-387-68918-0}

%Variations on ADMM \url{https://link.springer.com/article/10.1007/s10915-017-0621-6#Sec15}

\bibliography{ref}% Produces the bibliography via BibTeX.

%apsrev4-2.bst 2019-01-14 (MD) hand-edited version of apsrev4-1.bst
%Control: key (0)
%Control: author (8) initials jnrlst
%Control: editor formatted (1) identically to author
%Control: production of article title (0) allowed
%Control: page (0) single
%Control: year (1) truncated
%Control: production of eprint (0) enabled
\providecommand{\noopsort}[1]{}\providecommand{\singleletter}[1]{#1}%
\begin{thebibliography}{80}%
\makeatletter
\providecommand \@ifxundefined [1]{%
 \@ifx{#1\undefined}
}%
\providecommand \@ifnum [1]{%
 \ifnum #1\expandafter \@firstoftwo
 \else \expandafter \@secondoftwo
 \fi
}%
\providecommand \@ifx [1]{%
 \ifx #1\expandafter \@firstoftwo
 \else \expandafter \@secondoftwo
 \fi
}%
\providecommand \natexlab [1]{#1}%
\providecommand \enquote  [1]{``#1''}%
\providecommand \bibnamefont  [1]{#1}%
\providecommand \bibfnamefont [1]{#1}%
\providecommand \citenamefont [1]{#1}%
\providecommand \href@noop [0]{\@secondoftwo}%
\providecommand \href [0]{\begingroup \@sanitize@url \@href}%
\providecommand \@href[1]{\@@startlink{#1}\@@href}%
\providecommand \@@href[1]{\endgroup#1\@@endlink}%
\providecommand \@sanitize@url [0]{\catcode `\\12\catcode `\$12\catcode
  `\&12\catcode `\#12\catcode `\^12\catcode `\_12\catcode `\%12\relax}%
\providecommand \@@startlink[1]{}%
\providecommand \@@endlink[0]{}%
\providecommand \url  [0]{\begingroup\@sanitize@url \@url }%
\providecommand \@url [1]{\endgroup\@href {#1}{\urlprefix }}%
\providecommand \urlprefix  [0]{URL }%
\providecommand \Eprint [0]{\href }%
\providecommand \doibase [0]{https://doi.org/}%
\providecommand \selectlanguage [0]{\@gobble}%
\providecommand \bibinfo  [0]{\@secondoftwo}%
\providecommand \bibfield  [0]{\@secondoftwo}%
\providecommand \translation [1]{[#1]}%
\providecommand \BibitemOpen [0]{}%
\providecommand \bibitemStop [0]{}%
\providecommand \bibitemNoStop [0]{.\EOS\space}%
\providecommand \EOS [0]{\spacefactor3000\relax}%
\providecommand \BibitemShut  [1]{\csname bibitem#1\endcsname}%
\let\auto@bib@innerbib\@empty
%</preamble>
\bibitem [{\citenamefont {Kabanikhin}(2011)}]{Kabanikhin2011Dec}%
  \BibitemOpen
  \bibfield  {author} {\bibinfo {author} {\bibfnamefont {S.~I.}\ \bibnamefont
  {Kabanikhin}},\ }\href
  {https://www.degruyter.com/document/doi/10.1515/9783110224016/html?lang=en}
  {\emph {\bibinfo {title} {{Inverse and Ill-posed Problems}}}}\ (\bibinfo
  {publisher} {De Gruyter},\ \bibinfo {address} {Berlin, Germany},\ \bibinfo
  {year} {2011})\BibitemShut {NoStop}%
\bibitem [{\citenamefont {Vogel}\ and\ \citenamefont
  {Risken}(1989)}]{Vogel1989Sep}%
  \BibitemOpen
  \bibfield  {author} {\bibinfo {author} {\bibfnamefont {K.}~\bibnamefont
  {Vogel}}\ and\ \bibinfo {author} {\bibfnamefont {H.}~\bibnamefont {Risken}},\
  }\bibfield  {title} {\bibinfo {title} {{Determination of quasiprobability
  distributions in terms of probability distributions for the rotated
  quadrature phase}},\ }\href {https://doi.org/10.1103/PhysRevA.40.2847}
  {\bibfield  {journal} {\bibinfo  {journal} {Phys. Rev. A}\ }\textbf {\bibinfo
  {volume} {40}},\ \bibinfo {pages} {2847} (\bibinfo {year}
  {1989})}\BibitemShut {NoStop}%
\bibitem [{\citenamefont {Smithey}\ \emph {et~al.}(1993)\citenamefont
  {Smithey}, \citenamefont {Beck}, \citenamefont {Raymer},\ and\ \citenamefont
  {Faridani}}]{Smithey1993Mar}%
  \BibitemOpen
  \bibfield  {author} {\bibinfo {author} {\bibfnamefont {D.~T.}\ \bibnamefont
  {Smithey}}, \bibinfo {author} {\bibfnamefont {M.}~\bibnamefont {Beck}},
  \bibinfo {author} {\bibfnamefont {M.~G.}\ \bibnamefont {Raymer}},\ and\
  \bibinfo {author} {\bibfnamefont {A.}~\bibnamefont {Faridani}},\ }\bibfield
  {title} {\bibinfo {title} {{Measurement of the Wigner distribution and the
  density matrix of a light mode using optical homodyne tomography: Application
  to squeezed states and the vacuum}},\ }\href
  {https://doi.org/10.1103/PhysRevLett.70.1244} {\bibfield  {journal} {\bibinfo
   {journal} {Phys. Rev. Lett.}\ }\textbf {\bibinfo {volume} {70}},\ \bibinfo
  {pages} {1244} (\bibinfo {year} {1993})}\BibitemShut {NoStop}%
\bibitem [{\citenamefont {Lvovsky}\ and\ \citenamefont
  {Raymer}(2009)}]{Lvovsky2009Mar}%
  \BibitemOpen
  \bibfield  {author} {\bibinfo {author} {\bibfnamefont {A.~I.}\ \bibnamefont
  {Lvovsky}}\ and\ \bibinfo {author} {\bibfnamefont {M.~G.}\ \bibnamefont
  {Raymer}},\ }\bibfield  {title} {\bibinfo {title} {{Continuous-variable
  optical quantum-state tomography}},\ }\href
  {https://doi.org/10.1103/RevModPhys.81.299} {\bibfield  {journal} {\bibinfo
  {journal} {Rev. Mod. Phys.}\ }\textbf {\bibinfo {volume} {81}},\ \bibinfo
  {pages} {299} (\bibinfo {year} {2009})}\BibitemShut {NoStop}%
\bibitem [{\citenamefont {Leonhardt}\ \emph {et~al.}(1995)\citenamefont
  {Leonhardt}, \citenamefont {Paul},\ and\ \citenamefont
  {D{'}Ariano}}]{Leonhardt1995Dec}%
  \BibitemOpen
  \bibfield  {author} {\bibinfo {author} {\bibfnamefont {U.}~\bibnamefont
  {Leonhardt}}, \bibinfo {author} {\bibfnamefont {H.}~\bibnamefont {Paul}},\
  and\ \bibinfo {author} {\bibfnamefont {G.~M.}\ \bibnamefont {D{'}Ariano}},\
  }\bibfield  {title} {\bibinfo {title} {{Tomographic reconstruction of the
  density matrix via pattern functions}},\ }\href
  {https://doi.org/10.1103/PhysRevA.52.4899} {\bibfield  {journal} {\bibinfo
  {journal} {Phys. Rev. A}\ }\textbf {\bibinfo {volume} {52}},\ \bibinfo
  {pages} {4899} (\bibinfo {year} {1995})}\BibitemShut {NoStop}%
\bibitem [{\citenamefont {Richter}(1996)}]{Richter1996Mar}%
  \BibitemOpen
  \bibfield  {author} {\bibinfo {author} {\bibfnamefont {{\relax
  Th}.}~\bibnamefont {Richter}},\ }\bibfield  {title} {\bibinfo {title}
  {{Pattern functions used in tomographic reconstruction of photon statistics
  revisited}},\ }\href {https://doi.org/10.1016/0375-9601(96)00029-1}
  {\bibfield  {journal} {\bibinfo  {journal} {Phys. Lett. A}\ }\textbf
  {\bibinfo {volume} {211}},\ \bibinfo {pages} {327} (\bibinfo {year}
  {1996})}\BibitemShut {NoStop}%
\bibitem [{\citenamefont {Hradil}(1997)}]{Hradil1997Mar}%
  \BibitemOpen
  \bibfield  {author} {\bibinfo {author} {\bibfnamefont {Z.}~\bibnamefont
  {Hradil}},\ }\bibfield  {title} {\bibinfo {title} {{Quantum-state
  estimation}},\ }\href {https://doi.org/10.1103/PhysRevA.55.R1561} {\bibfield
  {journal} {\bibinfo  {journal} {Phys. Rev. A}\ }\textbf {\bibinfo {volume}
  {55}},\ \bibinfo {pages} {R1561} (\bibinfo {year} {1997})}\BibitemShut
  {NoStop}%
\bibitem [{\citenamefont {Banaszek}\ \emph
  {et~al.}(1999{\natexlab{a}})\citenamefont {Banaszek}, \citenamefont
  {D{'}Ariano}, \citenamefont {Paris},\ and\ \citenamefont
  {Sacchi}}]{Banaszek1999Dec}%
  \BibitemOpen
  \bibfield  {author} {\bibinfo {author} {\bibfnamefont {K.}~\bibnamefont
  {Banaszek}}, \bibinfo {author} {\bibfnamefont {G.~M.}\ \bibnamefont
  {D{'}Ariano}}, \bibinfo {author} {\bibfnamefont {M.~G.~A.}\ \bibnamefont
  {Paris}},\ and\ \bibinfo {author} {\bibfnamefont {M.~F.}\ \bibnamefont
  {Sacchi}},\ }\bibfield  {title} {\bibinfo {title} {{Maximum-likelihood
  estimation of the density matrix}},\ }\href
  {https://doi.org/10.1103/PhysRevA.61.010304} {\bibfield  {journal} {\bibinfo
  {journal} {Phys. Rev. A}\ }\textbf {\bibinfo {volume} {61}},\ \bibinfo
  {pages} {010304(R)} (\bibinfo {year} {1999}{\natexlab{a}})}\BibitemShut
  {NoStop}%
\bibitem [{\citenamefont {Lvovsky}(2004)}]{Lvovsky2004May}%
  \BibitemOpen
  \bibfield  {author} {\bibinfo {author} {\bibfnamefont {A.~I.}\ \bibnamefont
  {Lvovsky}},\ }\bibfield  {title} {\bibinfo {title} {{Iterative
  maximum-likelihood reconstruction in quantum homodyne tomography}},\ }\href
  {https://doi.org/10.1088/1464-4266/6/6/014} {\bibfield  {journal} {\bibinfo
  {journal} {J. Opt. B: Quantum Semiclassical Opt.}\ }\textbf {\bibinfo
  {volume} {6}},\ \bibinfo {pages} {S556} (\bibinfo {year} {2004})}\BibitemShut
  {NoStop}%
\bibitem [{\citenamefont {\ifmmode \check{R}\else
  \v{R}\fi{}eh\'a\ifmmode~\check{c}\else \v{c}\fi{}ek}\ \emph
  {et~al.}(2007)\citenamefont {\ifmmode \check{R}\else
  \v{R}\fi{}eh\'a\ifmmode~\check{c}\else \v{c}\fi{}ek}, \citenamefont {Hradil},
  \citenamefont {Knill},\ and\ \citenamefont {Lvovsky}}]{Rehacek2007Apr}%
  \BibitemOpen
  \bibfield  {author} {\bibinfo {author} {\bibfnamefont {J.}~\bibnamefont
  {\ifmmode \check{R}\else \v{R}\fi{}eh\'a\ifmmode~\check{c}\else
  \v{c}\fi{}ek}}, \bibinfo {author} {\bibfnamefont {Z.~c.~v.}\ \bibnamefont
  {Hradil}}, \bibinfo {author} {\bibfnamefont {E.}~\bibnamefont {Knill}},\ and\
  \bibinfo {author} {\bibfnamefont {A.~I.}\ \bibnamefont {Lvovsky}},\
  }\bibfield  {title} {\bibinfo {title} {{Diluted maximum-likelihood algorithm
  for quantum tomography}},\ }\href
  {https://doi.org/10.1103/PhysRevA.75.042108} {\bibfield  {journal} {\bibinfo
  {journal} {Phys. Rev. A}\ }\textbf {\bibinfo {volume} {75}},\ \bibinfo
  {pages} {042108} (\bibinfo {year} {2007})}\BibitemShut {NoStop}%
\bibitem [{\citenamefont {Shang}\ \emph {et~al.}(2017)\citenamefont {Shang},
  \citenamefont {Zhang},\ and\ \citenamefont {Ng}}]{Shang2017Jun}%
  \BibitemOpen
  \bibfield  {author} {\bibinfo {author} {\bibfnamefont {J.}~\bibnamefont
  {Shang}}, \bibinfo {author} {\bibfnamefont {Z.}~\bibnamefont {Zhang}},\ and\
  \bibinfo {author} {\bibfnamefont {H.~K.}\ \bibnamefont {Ng}},\ }\bibfield
  {title} {\bibinfo {title} {{Superfast maximum-likelihood reconstruction for
  quantum tomography}},\ }\href {https://doi.org/10.1103/PhysRevA.95.062336}
  {\bibfield  {journal} {\bibinfo  {journal} {Phys. Rev. A}\ }\textbf {\bibinfo
  {volume} {95}},\ \bibinfo {pages} {062336} (\bibinfo {year}
  {2017})}\BibitemShut {NoStop}%
\bibitem [{\citenamefont
  {Blume-Kohout}(2010{\natexlab{a}})}]{Blume-Kohout2010Nov}%
  \BibitemOpen
  \bibfield  {author} {\bibinfo {author} {\bibfnamefont {R.}~\bibnamefont
  {Blume-Kohout}},\ }\bibfield  {title} {\bibinfo {title} {{Hedged Maximum
  Likelihood Quantum State Estimation}},\ }\href
  {https://doi.org/10.1103/PhysRevLett.105.200504} {\bibfield  {journal}
  {\bibinfo  {journal} {Phys. Rev. Lett.}\ }\textbf {\bibinfo {volume} {105}},\
  \bibinfo {pages} {200504} (\bibinfo {year} {2010}{\natexlab{a}})}\BibitemShut
  {NoStop}%
\bibitem [{\citenamefont {Baumgratz}\ \emph {et~al.}(2013)\citenamefont
  {Baumgratz}, \citenamefont {N{\ifmmode\ddot{u}\else\"{u}\fi}{\ss}eler},
  \citenamefont {Cramer},\ and\ \citenamefont {Plenio}}]{Baumgratz2013Dec}%
  \BibitemOpen
  \bibfield  {author} {\bibinfo {author} {\bibfnamefont {T.}~\bibnamefont
  {Baumgratz}}, \bibinfo {author} {\bibfnamefont {A.}~\bibnamefont
  {N{\ifmmode\ddot{u}\else\"{u}\fi}{\ss}eler}}, \bibinfo {author}
  {\bibfnamefont {M.}~\bibnamefont {Cramer}},\ and\ \bibinfo {author}
  {\bibfnamefont {M.~B.}\ \bibnamefont {Plenio}},\ }\bibfield  {title}
  {\bibinfo {title} {{A scalable maximum likelihood method for quantum state
  tomography}},\ }\href {https://doi.org/10.1088/1367-2630/15/12/125004}
  {\bibfield  {journal} {\bibinfo  {journal} {New J. Phys.}\ }\textbf {\bibinfo
  {volume} {15}},\ \bibinfo {pages} {125004} (\bibinfo {year}
  {2013})}\BibitemShut {NoStop}%
\bibitem [{\citenamefont
  {Blume-Kohout}(2010{\natexlab{b}})}]{Blume-Kohout2010Apr}%
  \BibitemOpen
  \bibfield  {author} {\bibinfo {author} {\bibfnamefont {R.}~\bibnamefont
  {Blume-Kohout}},\ }\bibfield  {title} {\bibinfo {title} {{Optimal, reliable
  estimation of quantum states}},\ }\href
  {https://doi.org/10.1088/1367-2630/12/4/043034} {\bibfield  {journal}
  {\bibinfo  {journal} {New J. Phys.}\ }\textbf {\bibinfo {volume} {12}},\
  \bibinfo {pages} {043034} (\bibinfo {year} {2010}{\natexlab{b}})}\BibitemShut
  {NoStop}%
\bibitem [{\citenamefont {Granade}\ \emph {et~al.}(2016)\citenamefont
  {Granade}, \citenamefont {Combes},\ and\ \citenamefont
  {Cory}}]{Granade2016Mar}%
  \BibitemOpen
  \bibfield  {author} {\bibinfo {author} {\bibfnamefont {C.}~\bibnamefont
  {Granade}}, \bibinfo {author} {\bibfnamefont {J.}~\bibnamefont {Combes}},\
  and\ \bibinfo {author} {\bibfnamefont {D.~G.}\ \bibnamefont {Cory}},\
  }\bibfield  {title} {\bibinfo {title} {{Practical Bayesian tomography}},\
  }\href {https://doi.org/10.1088/1367-2630/18/3/033024} {\bibfield  {journal}
  {\bibinfo  {journal} {New J. Phys.}\ }\textbf {\bibinfo {volume} {18}},\
  \bibinfo {pages} {033024} (\bibinfo {year} {2016})}\BibitemShut {NoStop}%
\bibitem [{\citenamefont {Lukens}\ \emph {et~al.}(2020)\citenamefont {Lukens},
  \citenamefont {Law}, \citenamefont {Jasra},\ and\ \citenamefont
  {Lougovski}}]{Lukens2020Jun}%
  \BibitemOpen
  \bibfield  {author} {\bibinfo {author} {\bibfnamefont {J.~M.}\ \bibnamefont
  {Lukens}}, \bibinfo {author} {\bibfnamefont {K.~J.~H.}\ \bibnamefont {Law}},
  \bibinfo {author} {\bibfnamefont {A.}~\bibnamefont {Jasra}},\ and\ \bibinfo
  {author} {\bibfnamefont {P.}~\bibnamefont {Lougovski}},\ }\bibfield  {title}
  {\bibinfo {title} {{A practical and efficient approach for Bayesian quantum
  state estimation}},\ }\href {https://doi.org/10.1088/1367-2630/ab8efa}
  {\bibfield  {journal} {\bibinfo  {journal} {New J. Phys.}\ }\textbf {\bibinfo
  {volume} {22}},\ \bibinfo {pages} {063038} (\bibinfo {year}
  {2020})}\BibitemShut {NoStop}%
\bibitem [{\citenamefont {Tiunov}\ \emph {et~al.}(2020)\citenamefont {Tiunov},
  \citenamefont {Tiunova~(Vyborova)}, \citenamefont {Ulanov}, \citenamefont
  {Lvovsky},\ and\ \citenamefont {Fedorov}}]{Tiunov2020May}%
  \BibitemOpen
  \bibfield  {author} {\bibinfo {author} {\bibfnamefont {E.~S.}\ \bibnamefont
  {Tiunov}}, \bibinfo {author} {\bibfnamefont {V.~V.}\ \bibnamefont
  {Tiunova~(Vyborova)}}, \bibinfo {author} {\bibfnamefont {A.~E.}\ \bibnamefont
  {Ulanov}}, \bibinfo {author} {\bibfnamefont {A.~I.}\ \bibnamefont
  {Lvovsky}},\ and\ \bibinfo {author} {\bibfnamefont {A.~K.}\ \bibnamefont
  {Fedorov}},\ }\bibfield  {title} {\bibinfo {title} {{Experimental quantum
  homodyne tomography via machine learning}},\ }\href
  {https://doi.org/10.1364/OPTICA.389482} {\bibfield  {journal} {\bibinfo
  {journal} {Optica}\ }\textbf {\bibinfo {volume} {7}},\ \bibinfo {pages} {448}
  (\bibinfo {year} {2020})}\BibitemShut {NoStop}%
\bibitem [{\citenamefont {Ghosh}\ \emph {et~al.}(2020)\citenamefont {Ghosh},
  \citenamefont {Opala}, \citenamefont {Matuszewski}, \citenamefont {Paterek},\
  and\ \citenamefont {Liew}}]{Ghosh2020Jul}%
  \BibitemOpen
  \bibfield  {author} {\bibinfo {author} {\bibfnamefont {S.}~\bibnamefont
  {Ghosh}}, \bibinfo {author} {\bibfnamefont {A.}~\bibnamefont {Opala}},
  \bibinfo {author} {\bibfnamefont {M.}~\bibnamefont {Matuszewski}}, \bibinfo
  {author} {\bibfnamefont {T.}~\bibnamefont {Paterek}},\ and\ \bibinfo {author}
  {\bibfnamefont {T.~C.~H.}\ \bibnamefont {Liew}},\ }\bibfield  {title}
  {\bibinfo {title} {{Reconstructing Quantum States With Quantum Reservoir
  Networks}},\ }\href {https://doi.org/10.1109/TNNLS.2020.3009716} {\bibfield
  {journal} {\bibinfo  {journal} {IEEE Trans. Neural Networks Learn. Syst.}\
  }\textbf {\bibinfo {volume} {32}},\ \bibinfo {pages} {3148} (\bibinfo {year}
  {2020})}\BibitemShut {NoStop}%
\bibitem [{\citenamefont {Ahmed}\ \emph
  {et~al.}(2021{\natexlab{a}})\citenamefont {Ahmed}, \citenamefont {S\'{a}nchez
  Mu\~noz}, \citenamefont {Nori},\ and\ \citenamefont
  {Kockum}}]{Ahmed2021Sep1}%
  \BibitemOpen
  \bibfield  {author} {\bibinfo {author} {\bibfnamefont {S.}~\bibnamefont
  {Ahmed}}, \bibinfo {author} {\bibfnamefont {C.}~\bibnamefont {S\'{a}nchez
  Mu\~noz}}, \bibinfo {author} {\bibfnamefont {F.}~\bibnamefont {Nori}},\ and\
  \bibinfo {author} {\bibfnamefont {A.~F.}\ \bibnamefont {Kockum}},\ }\bibfield
   {title} {\bibinfo {title} {{Quantum State Tomography with Conditional
  Generative Adversarial Networks}},\ }\href
  {https://doi.org/10.1103/PhysRevLett.127.140502} {\bibfield  {journal}
  {\bibinfo  {journal} {Phys. Rev. Lett.}\ }\textbf {\bibinfo {volume} {127}},\
  \bibinfo {pages} {140502} (\bibinfo {year} {2021}{\natexlab{a}})}\BibitemShut
  {NoStop}%
\bibitem [{\citenamefont {Ahmed}\ \emph
  {et~al.}(2021{\natexlab{b}})\citenamefont {Ahmed}, \citenamefont {S\'{a}nchez
  Mu\~noz}, \citenamefont {Nori},\ and\ \citenamefont {Kockum}}]{Ahmed2021Sep}%
  \BibitemOpen
  \bibfield  {author} {\bibinfo {author} {\bibfnamefont {S.}~\bibnamefont
  {Ahmed}}, \bibinfo {author} {\bibfnamefont {C.}~\bibnamefont {S\'{a}nchez
  Mu\~noz}}, \bibinfo {author} {\bibfnamefont {F.}~\bibnamefont {Nori}},\ and\
  \bibinfo {author} {\bibfnamefont {A.~F.}\ \bibnamefont {Kockum}},\ }\bibfield
   {title} {\bibinfo {title} {{Classification and reconstruction of optical
  quantum states with deep neural networks}},\ }\href
  {https://doi.org/10.1103/PhysRevResearch.3.033278} {\bibfield  {journal}
  {\bibinfo  {journal} {Phys. Rev. Res.}\ }\textbf {\bibinfo {volume} {3}},\
  \bibinfo {pages} {033278} (\bibinfo {year} {2021}{\natexlab{b}})}\BibitemShut
  {NoStop}%
\bibitem [{\citenamefont {Chuang}\ and\ \citenamefont
  {Nielsen}(1997)}]{Chuang1997Nov}%
  \BibitemOpen
  \bibfield  {author} {\bibinfo {author} {\bibfnamefont {I.~L.}\ \bibnamefont
  {Chuang}}\ and\ \bibinfo {author} {\bibfnamefont {M.~A.}\ \bibnamefont
  {Nielsen}},\ }\bibfield  {title} {\bibinfo {title} {{Prescription for
  experimental determination of the dynamics of a quantum black box}},\ }\href
  {https://doi.org/10.1080/09500349708231894} {\bibfield  {journal} {\bibinfo
  {journal} {J. Mod. Opt.}\ }\textbf {\bibinfo {volume} {44}},\ \bibinfo
  {pages} {2455} (\bibinfo {year} {1997})}\BibitemShut {NoStop}%
\bibitem [{\citenamefont {Bialczak}\ \emph {et~al.}(2010)\citenamefont
  {Bialczak}, \citenamefont {Ansmann}, \citenamefont {Hofheinz}, \citenamefont
  {Lucero}, \citenamefont {Neeley}, \citenamefont {O{'}Connell}, \citenamefont
  {Sank}, \citenamefont {Wang}, \citenamefont {Wenner}, \citenamefont
  {Steffen}, \citenamefont {Cleland},\ and\ \citenamefont
  {Martinis}}]{Bialczak2010Jun}%
  \BibitemOpen
  \bibfield  {author} {\bibinfo {author} {\bibfnamefont {R.~C.}\ \bibnamefont
  {Bialczak}}, \bibinfo {author} {\bibfnamefont {M.}~\bibnamefont {Ansmann}},
  \bibinfo {author} {\bibfnamefont {M.}~\bibnamefont {Hofheinz}}, \bibinfo
  {author} {\bibfnamefont {E.}~\bibnamefont {Lucero}}, \bibinfo {author}
  {\bibfnamefont {M.}~\bibnamefont {Neeley}}, \bibinfo {author} {\bibfnamefont
  {A.~D.}\ \bibnamefont {O{'}Connell}}, \bibinfo {author} {\bibfnamefont
  {D.}~\bibnamefont {Sank}}, \bibinfo {author} {\bibfnamefont {H.}~\bibnamefont
  {Wang}}, \bibinfo {author} {\bibfnamefont {J.}~\bibnamefont {Wenner}},
  \bibinfo {author} {\bibfnamefont {M.}~\bibnamefont {Steffen}}, \bibinfo
  {author} {\bibfnamefont {A.~N.}\ \bibnamefont {Cleland}},\ and\ \bibinfo
  {author} {\bibfnamefont {J.~M.}\ \bibnamefont {Martinis}},\ }\bibfield
  {title} {\bibinfo {title} {{Quantum process tomography of a universal
  entangling gate implemented with Josephson phase qubits}},\ }\href
  {https://doi.org/10.1038/nphys1639} {\bibfield  {journal} {\bibinfo
  {journal} {Nat. Phys.}\ }\textbf {\bibinfo {volume} {6}},\ \bibinfo {pages}
  {409} (\bibinfo {year} {2010})}\BibitemShut {NoStop}%
\bibitem [{\citenamefont {Altepeter}\ \emph {et~al.}(2003)\citenamefont
  {Altepeter}, \citenamefont {Branning}, \citenamefont {Jeffrey}, \citenamefont
  {Wei}, \citenamefont {Kwiat}, \citenamefont {Thew}, \citenamefont
  {O{'}Brien}, \citenamefont {Nielsen},\ and\ \citenamefont
  {White}}]{Altepeter2003May}%
  \BibitemOpen
  \bibfield  {author} {\bibinfo {author} {\bibfnamefont {J.~B.}\ \bibnamefont
  {Altepeter}}, \bibinfo {author} {\bibfnamefont {D.}~\bibnamefont {Branning}},
  \bibinfo {author} {\bibfnamefont {E.}~\bibnamefont {Jeffrey}}, \bibinfo
  {author} {\bibfnamefont {T.~C.}\ \bibnamefont {Wei}}, \bibinfo {author}
  {\bibfnamefont {P.~G.}\ \bibnamefont {Kwiat}}, \bibinfo {author}
  {\bibfnamefont {R.~T.}\ \bibnamefont {Thew}}, \bibinfo {author}
  {\bibfnamefont {J.~L.}\ \bibnamefont {O{'}Brien}}, \bibinfo {author}
  {\bibfnamefont {M.~A.}\ \bibnamefont {Nielsen}},\ and\ \bibinfo {author}
  {\bibfnamefont {A.~G.}\ \bibnamefont {White}},\ }\bibfield  {title} {\bibinfo
  {title} {{Ancilla-Assisted Quantum Process Tomography}},\ }\href
  {https://doi.org/10.1103/PhysRevLett.90.193601} {\bibfield  {journal}
  {\bibinfo  {journal} {Phys. Rev. Lett.}\ }\textbf {\bibinfo {volume} {90}},\
  \bibinfo {pages} {193601} (\bibinfo {year} {2003})}\BibitemShut {NoStop}%
\bibitem [{\citenamefont {Mohseni}\ \emph {et~al.}(2008)\citenamefont
  {Mohseni}, \citenamefont {Rezakhani},\ and\ \citenamefont
  {Lidar}}]{Mohseni2008Mar}%
  \BibitemOpen
  \bibfield  {author} {\bibinfo {author} {\bibfnamefont {M.}~\bibnamefont
  {Mohseni}}, \bibinfo {author} {\bibfnamefont {A.~T.}\ \bibnamefont
  {Rezakhani}},\ and\ \bibinfo {author} {\bibfnamefont {D.~A.}\ \bibnamefont
  {Lidar}},\ }\bibfield  {title} {\bibinfo {title} {{Quantum-process
  tomography: Resource analysis of different strategies}},\ }\href
  {https://doi.org/10.1103/PhysRevA.77.032322} {\bibfield  {journal} {\bibinfo
  {journal} {Phys. Rev. A}\ }\textbf {\bibinfo {volume} {77}},\ \bibinfo
  {pages} {032322} (\bibinfo {year} {2008})}\BibitemShut {NoStop}%
\bibitem [{\citenamefont {Rahimi-Keshari}\ \emph {et~al.}(2011)\citenamefont
  {Rahimi-Keshari}, \citenamefont {Scherer}, \citenamefont {Mann},
  \citenamefont {Rezakhani}, \citenamefont {Lvovsky},\ and\ \citenamefont
  {Sanders}}]{Rahimi-Keshari2011Jan}%
  \BibitemOpen
  \bibfield  {author} {\bibinfo {author} {\bibfnamefont {S.}~\bibnamefont
  {Rahimi-Keshari}}, \bibinfo {author} {\bibfnamefont {A.}~\bibnamefont
  {Scherer}}, \bibinfo {author} {\bibfnamefont {A.}~\bibnamefont {Mann}},
  \bibinfo {author} {\bibfnamefont {A.~T.}\ \bibnamefont {Rezakhani}}, \bibinfo
  {author} {\bibfnamefont {A.~I.}\ \bibnamefont {Lvovsky}},\ and\ \bibinfo
  {author} {\bibfnamefont {B.~C.}\ \bibnamefont {Sanders}},\ }\bibfield
  {title} {\bibinfo {title} {{Quantum process tomography with coherent
  states}},\ }\href {https://doi.org/10.1088/1367-2630/13/1/013006} {\bibfield
  {journal} {\bibinfo  {journal} {New J. Phys.}\ }\textbf {\bibinfo {volume}
  {13}},\ \bibinfo {pages} {013006} (\bibinfo {year} {2011})}\BibitemShut
  {NoStop}%
\bibitem [{\citenamefont {Ghalaii}\ and\ \citenamefont
  {Rezakhani}(2017)}]{Ghalaii2017Mar}%
  \BibitemOpen
  \bibfield  {author} {\bibinfo {author} {\bibfnamefont {M.}~\bibnamefont
  {Ghalaii}}\ and\ \bibinfo {author} {\bibfnamefont {A.~T.}\ \bibnamefont
  {Rezakhani}},\ }\bibfield  {title} {\bibinfo {title} {{Scheme for
  coherent-state quantum process tomography via normally-ordered moments}},\
  }\href {https://doi.org/10.1103/PhysRevA.95.032336} {\bibfield  {journal}
  {\bibinfo  {journal} {Phys. Rev. A}\ }\textbf {\bibinfo {volume} {95}},\
  \bibinfo {pages} {032336} (\bibinfo {year} {2017})}\BibitemShut {NoStop}%
\bibitem [{\citenamefont {Boyd}\ and\ \citenamefont
  {Vandenberghe}(2004)}]{Boyd2004Mar}%
  \BibitemOpen
  \bibfield  {author} {\bibinfo {author} {\bibfnamefont {S.}~\bibnamefont
  {Boyd}}\ and\ \bibinfo {author} {\bibfnamefont {L.}~\bibnamefont
  {Vandenberghe}},\ }\href {https://web.stanford.edu/~boyd/cvxbook} {\emph
  {\bibinfo {title} {{Convex Optimization}}}}\ (\bibinfo  {publisher}
  {Cambridge University Press},\ \bibinfo {address} {Cambridge, England, UK},\
  \bibinfo {year} {2004})\BibitemShut {NoStop}%
\bibitem [{\citenamefont {Huang}\ \emph {et~al.}(2020)\citenamefont {Huang},
  \citenamefont {Gao}, \citenamefont {Jiao}, \citenamefont {Yan}, \citenamefont
  {Zhang}, \citenamefont {Chen}, \citenamefont {Zhang}, \citenamefont {Ji},\
  and\ \citenamefont {Jin}}]{Huang2020Feb}%
  \BibitemOpen
  \bibfield  {author} {\bibinfo {author} {\bibfnamefont {X.-L.}\ \bibnamefont
  {Huang}}, \bibinfo {author} {\bibfnamefont {J.}~\bibnamefont {Gao}}, \bibinfo
  {author} {\bibfnamefont {Z.-Q.}\ \bibnamefont {Jiao}}, \bibinfo {author}
  {\bibfnamefont {Z.-Q.}\ \bibnamefont {Yan}}, \bibinfo {author} {\bibfnamefont
  {Z.-Y.}\ \bibnamefont {Zhang}}, \bibinfo {author} {\bibfnamefont {D.-Y.}\
  \bibnamefont {Chen}}, \bibinfo {author} {\bibfnamefont {X.}~\bibnamefont
  {Zhang}}, \bibinfo {author} {\bibfnamefont {L.}~\bibnamefont {Ji}},\ and\
  \bibinfo {author} {\bibfnamefont {X.-M.}\ \bibnamefont {Jin}},\ }\bibfield
  {title} {\bibinfo {title} {{Reconstruction of quantum channel via convex
  optimization}},\ }\href {https://doi.org/10.1016/j.scib.2019.11.009}
  {\bibfield  {journal} {\bibinfo  {journal} {Science Bulletin}\ }\textbf
  {\bibinfo {volume} {65}},\ \bibinfo {pages} {286} (\bibinfo {year}
  {2020})}\BibitemShut {NoStop}%
\bibitem [{\citenamefont {Kim}\ \emph {et~al.}(2021)\citenamefont {Kim},
  \citenamefont {Kollias}, \citenamefont {Kalev}, \citenamefont {Wei},\ and\
  \citenamefont {Kyrillidis}}]{Kim2021Apr}%
  \BibitemOpen
  \bibfield  {author} {\bibinfo {author} {\bibfnamefont {J.~L.}\ \bibnamefont
  {Kim}}, \bibinfo {author} {\bibfnamefont {G.}~\bibnamefont {Kollias}},
  \bibinfo {author} {\bibfnamefont {A.}~\bibnamefont {Kalev}}, \bibinfo
  {author} {\bibfnamefont {K.~X.}\ \bibnamefont {Wei}},\ and\ \bibinfo {author}
  {\bibfnamefont {A.}~\bibnamefont {Kyrillidis}},\ }\bibfield  {title}
  {\bibinfo {title} {{Fast quantum state reconstruction via accelerated
  non-convex programming}},\ }\href {https://arxiv.org/abs/2104.07006v3}
  {\bibfield  {journal} {\bibinfo  {journal} {arXiv}\ } (\bibinfo {year}
  {2021})},\ \Eprint {https://arxiv.org/abs/2104.07006} {2104.07006}
  \BibitemShut {NoStop}%
\bibitem [{\citenamefont {Zhang}\ \emph {et~al.}(2012)\citenamefont {Zhang},
  \citenamefont {Coldenstrodt-Ronge}, \citenamefont {Datta}, \citenamefont
  {Puentes}, \citenamefont {Lundeen}, \citenamefont {Jin}, \citenamefont
  {Smith}, \citenamefont {Plenio},\ and\ \citenamefont
  {Walmsley}}]{Zhang2012Jun}%
  \BibitemOpen
  \bibfield  {author} {\bibinfo {author} {\bibfnamefont {L.}~\bibnamefont
  {Zhang}}, \bibinfo {author} {\bibfnamefont {H.~B.}\ \bibnamefont
  {Coldenstrodt-Ronge}}, \bibinfo {author} {\bibfnamefont {A.}~\bibnamefont
  {Datta}}, \bibinfo {author} {\bibfnamefont {G.}~\bibnamefont {Puentes}},
  \bibinfo {author} {\bibfnamefont {J.~S.}\ \bibnamefont {Lundeen}}, \bibinfo
  {author} {\bibfnamefont {X.-M.}\ \bibnamefont {Jin}}, \bibinfo {author}
  {\bibfnamefont {B.~J.}\ \bibnamefont {Smith}}, \bibinfo {author}
  {\bibfnamefont {M.~B.}\ \bibnamefont {Plenio}},\ and\ \bibinfo {author}
  {\bibfnamefont {I.~A.}\ \bibnamefont {Walmsley}},\ }\bibfield  {title}
  {\bibinfo {title} {{Mapping coherence in measurement via full quantum
  tomography of a hybrid optical detector - Nature Photonics}},\ }\href
  {https://doi.org/10.1038/nphoton.2012.107} {\bibfield  {journal} {\bibinfo
  {journal} {Nat. Photonics}\ }\textbf {\bibinfo {volume} {6}},\ \bibinfo
  {pages} {364} (\bibinfo {year} {2012})}\BibitemShut {NoStop}%
\bibitem [{\citenamefont {Cooper}\ \emph {et~al.}(2014)\citenamefont {Cooper},
  \citenamefont {Karpi{\ifmmode\acute{n}\else\'{n}\fi}ski},\ and\ \citenamefont
  {Smith}}]{Cooper2014Jul}%
  \BibitemOpen
  \bibfield  {author} {\bibinfo {author} {\bibfnamefont {M.}~\bibnamefont
  {Cooper}}, \bibinfo {author} {\bibfnamefont {M.}~\bibnamefont
  {Karpi{\ifmmode\acute{n}\else\'{n}\fi}ski}},\ and\ \bibinfo {author}
  {\bibfnamefont {B.~J.}\ \bibnamefont {Smith}},\ }\bibfield  {title} {\bibinfo
  {title} {{Local mapping of detector response for reliable quantum state
  estimation - Nature Communications}},\ }\href
  {https://doi.org/10.1038/ncomms5332} {\bibfield  {journal} {\bibinfo
  {journal} {Nat. Commun.}\ }\textbf {\bibinfo {volume} {5}},\ \bibinfo {pages}
  {1} (\bibinfo {year} {2014})}\BibitemShut {NoStop}%
\bibitem [{\citenamefont {Riofr{\ifmmode\acute{\imath}\else\'{\i}\fi}o}\ \emph
  {et~al.}(2017)\citenamefont {Riofr{\ifmmode\acute{\imath}\else\'{\i}\fi}o},
  \citenamefont {Gross}, \citenamefont {Flammia}, \citenamefont {Monz},
  \citenamefont {Nigg}, \citenamefont {Blatt},\ and\ \citenamefont
  {Eisert}}]{Riofrio2017May}%
  \BibitemOpen
  \bibfield  {author} {\bibinfo {author} {\bibfnamefont {C.~A.}\ \bibnamefont
  {Riofr{\ifmmode\acute{\imath}\else\'{\i}\fi}o}}, \bibinfo {author}
  {\bibfnamefont {D.}~\bibnamefont {Gross}}, \bibinfo {author} {\bibfnamefont
  {S.~T.}\ \bibnamefont {Flammia}}, \bibinfo {author} {\bibfnamefont
  {T.}~\bibnamefont {Monz}}, \bibinfo {author} {\bibfnamefont {D.}~\bibnamefont
  {Nigg}}, \bibinfo {author} {\bibfnamefont {R.}~\bibnamefont {Blatt}},\ and\
  \bibinfo {author} {\bibfnamefont {J.}~\bibnamefont {Eisert}},\ }\bibfield
  {title} {\bibinfo {title} {{Experimental quantum compressed sensing for a
  seven-qubit system - Nature Communications}},\ }\href
  {https://doi.org/10.1038/ncomms15305} {\bibfield  {journal} {\bibinfo
  {journal} {Nat. Commun.}\ }\textbf {\bibinfo {volume} {8}},\ \bibinfo {pages}
  {1} (\bibinfo {year} {2017})}\BibitemShut {NoStop}%
\bibitem [{\citenamefont {Flammia}\ \emph {et~al.}(2012)\citenamefont
  {Flammia}, \citenamefont {Gross}, \citenamefont {Liu},\ and\ \citenamefont
  {Eisert}}]{Flammia2012Sep}%
  \BibitemOpen
  \bibfield  {author} {\bibinfo {author} {\bibfnamefont {S.~T.}\ \bibnamefont
  {Flammia}}, \bibinfo {author} {\bibfnamefont {D.}~\bibnamefont {Gross}},
  \bibinfo {author} {\bibfnamefont {Y.-K.}\ \bibnamefont {Liu}},\ and\ \bibinfo
  {author} {\bibfnamefont {J.}~\bibnamefont {Eisert}},\ }\bibfield  {title}
  {\bibinfo {title} {{Quantum tomography via compressed sensing: error bounds,
  sample complexity and efficient estimators}},\ }\href
  {https://doi.org/10.1088/1367-2630/14/9/095022} {\bibfield  {journal}
  {\bibinfo  {journal} {New J. Phys.}\ }\textbf {\bibinfo {volume} {14}},\
  \bibinfo {pages} {095022} (\bibinfo {year} {2012})}\BibitemShut {NoStop}%
\bibitem [{\citenamefont {Gross}\ \emph {et~al.}(2010)\citenamefont {Gross},
  \citenamefont {Liu}, \citenamefont {Flammia}, \citenamefont {Becker},\ and\
  \citenamefont {Eisert}}]{Gross2010Oct}%
  \BibitemOpen
  \bibfield  {author} {\bibinfo {author} {\bibfnamefont {D.}~\bibnamefont
  {Gross}}, \bibinfo {author} {\bibfnamefont {Y.-K.}\ \bibnamefont {Liu}},
  \bibinfo {author} {\bibfnamefont {S.~T.}\ \bibnamefont {Flammia}}, \bibinfo
  {author} {\bibfnamefont {S.}~\bibnamefont {Becker}},\ and\ \bibinfo {author}
  {\bibfnamefont {J.}~\bibnamefont {Eisert}},\ }\bibfield  {title} {\bibinfo
  {title} {{Quantum State Tomography via Compressed Sensing}},\ }\href
  {https://doi.org/10.1103/PhysRevLett.105.150401} {\bibfield  {journal}
  {\bibinfo  {journal} {Phys. Rev. Lett.}\ }\textbf {\bibinfo {volume} {105}},\
  \bibinfo {pages} {150401} (\bibinfo {year} {2010})}\BibitemShut {NoStop}%
\bibitem [{\citenamefont {Smith}\ \emph {et~al.}(2013)\citenamefont {Smith},
  \citenamefont {Riofr\'{\i}o}, \citenamefont {Anderson}, \citenamefont
  {Sosa-Martinez}, \citenamefont {Deutsch},\ and\ \citenamefont
  {Jessen}}]{Smith2013Mar}%
  \BibitemOpen
  \bibfield  {author} {\bibinfo {author} {\bibfnamefont {A.}~\bibnamefont
  {Smith}}, \bibinfo {author} {\bibfnamefont {C.~A.}\ \bibnamefont
  {Riofr\'{\i}o}}, \bibinfo {author} {\bibfnamefont {B.~E.}\ \bibnamefont
  {Anderson}}, \bibinfo {author} {\bibfnamefont {H.}~\bibnamefont
  {Sosa-Martinez}}, \bibinfo {author} {\bibfnamefont {I.~H.}\ \bibnamefont
  {Deutsch}},\ and\ \bibinfo {author} {\bibfnamefont {P.~S.}\ \bibnamefont
  {Jessen}},\ }\bibfield  {title} {\bibinfo {title} {{Quantum state tomography
  by continuous measurement and compressed sensing}},\ }\href
  {https://doi.org/10.1103/PhysRevA.87.030102} {\bibfield  {journal} {\bibinfo
  {journal} {Phys. Rev. A}\ }\textbf {\bibinfo {volume} {87}},\ \bibinfo
  {pages} {030102(R)} (\bibinfo {year} {2013})}\BibitemShut {NoStop}%
\bibitem [{Note1()}]{Note1}%
  \BibitemOpen
  \bibinfo {note} {If one does not have a priori information about the maximum
  photon population, to ensure there is no error due to the cutoff one can
  increase it until the result does not change with increasing
  dimension.}\BibitemShut {Stop}%
\bibitem [{\citenamefont {Nielsen}\ and\ \citenamefont
  {Chuang}(2010)}]{Nielsen2010-fe}%
  \BibitemOpen
  \bibfield  {author} {\bibinfo {author} {\bibfnamefont {M.~A.}\ \bibnamefont
  {Nielsen}}\ and\ \bibinfo {author} {\bibfnamefont {I.~L.}\ \bibnamefont
  {Chuang}},\ }\href@noop {} {\emph {\bibinfo {title} {Quantum Computation and
  Quantum Information}}}\ (\bibinfo  {publisher} {Cambridge University Press},\
  \bibinfo {address} {Cambridge, England},\ \bibinfo {year} {2010})\BibitemShut
  {NoStop}%
\bibitem [{Note2()}]{Note2}%
  \BibitemOpen
  \bibinfo {note} {In general, $k$ can be continuous. Nevertheless, one can
  always construct POVM elements labeled by a discrete index by binning the
  measurement results. This will be done for homodyne data in section~\ref
  {sec:homodyne}.}\BibitemShut {Stop}%
\bibitem [{\citenamefont {Peres}(2002)}]{Peres2002}%
  \BibitemOpen
  \bibfield  {author} {\bibinfo {author} {\bibfnamefont {A.}~\bibnamefont
  {Peres}},\ }\href {https://doi.org/10.1007/0-306-47120-5} {\emph {\bibinfo
  {title} {{Quantum Theory: Concepts and Methods}}}}\ (\bibinfo  {publisher}
  {Springer},\ \bibinfo {address} {Dordrecht, The Netherlands},\ \bibinfo
  {year} {2002})\BibitemShut {NoStop}%
\bibitem [{Note3()}]{Note3}%
  \BibitemOpen
  \bibinfo {note} {The notation $\protect \vec b$ is chosen instead of
  $\protect \vec p$ to avoid confusion with the vectorized density matrix
  $\protect \vec \rho $}\BibitemShut {NoStop}%
\bibitem [{\citenamefont {Mogilevtsev}\ \emph {et~al.}(1997)\citenamefont
  {Mogilevtsev}, \citenamefont {Hradil},\ and\ \citenamefont
  {Pe{\ifmmode\check{r}\else\v{r}\fi}ina}}]{Mogilevtsev1997Nov}%
  \BibitemOpen
  \bibfield  {author} {\bibinfo {author} {\bibfnamefont {D.}~\bibnamefont
  {Mogilevtsev}}, \bibinfo {author} {\bibfnamefont {Z.}~\bibnamefont
  {Hradil}},\ and\ \bibinfo {author} {\bibfnamefont {J.}~\bibnamefont
  {Pe{\ifmmode\check{r}\else\v{r}\fi}ina}},\ }\bibfield  {title} {\bibinfo
  {title} {{Homodyne reconstruction of density matrix in fock-state basis:
  Deterministic versus maximum likelihood approach}},\ }\href
  {https://doi.org/10.1080/09500349708231882} {\bibfield  {journal} {\bibinfo
  {journal} {J. Mod. Opt.}\ }\textbf {\bibinfo {volume} {44}},\ \bibinfo
  {pages} {2261} (\bibinfo {year} {1997})}\BibitemShut {NoStop}%
\bibitem [{Note4()}]{Note4}%
  \BibitemOpen
  \bibinfo {note} {For a vector $x \in C^{m}$, the $\ell _{2}$ norm is defined
  as $\|x\|_{\ell _{2}}=$ $\protect \sqrt {x^{\dagger } x}=\protect \sqrt
  {\DOTSB \sum@ \slimits@ _{i=1}^{m}\left |x_{i}\right |^{2}}$.}\BibitemShut
  {Stop}%
\bibitem [{\citenamefont {Boyd}\ and\ \citenamefont
  {Vandenberghe}(2018)}]{Boyd2018-ad}%
  \BibitemOpen
  \bibfield  {author} {\bibinfo {author} {\bibfnamefont {S.}~\bibnamefont
  {Boyd}}\ and\ \bibinfo {author} {\bibfnamefont {L.}~\bibnamefont
  {Vandenberghe}},\ }\href {https://web.stanford.edu/~boyd/vmls} {\emph
  {\bibinfo {title} {Introduction to applied linear algebra -- Vectors,
  Matrices, and Least Squares}}}\ (\bibinfo  {publisher} {Cambridge University
  Press},\ \bibinfo {address} {Cambridge, England},\ \bibinfo {year}
  {2018})\BibitemShut {NoStop}%
\bibitem [{\citenamefont {Kress}(1998)}]{Kress1998}%
  \BibitemOpen
  \bibfield  {author} {\bibinfo {author} {\bibfnamefont {R.}~\bibnamefont
  {Kress}},\ }\href {https://doi.org/10.1007/978-1-4612-0599-9} {\emph
  {\bibinfo {title} {{Numerical Analysis}}}}\ (\bibinfo  {publisher} {Springer,
  New York, NY},\ \bibinfo {address} {New York, NY, USA},\ \bibinfo {year}
  {1998})\BibitemShut {NoStop}%
\bibitem [{\citenamefont {Engl}\ \emph {et~al.}(1996)\citenamefont {Engl},
  \citenamefont {Hanke},\ and\ \citenamefont {Neubauer}}]{Engl1996-wx}%
  \BibitemOpen
  \bibfield  {author} {\bibinfo {author} {\bibfnamefont {H.~W.}\ \bibnamefont
  {Engl}}, \bibinfo {author} {\bibfnamefont {M.}~\bibnamefont {Hanke}},\ and\
  \bibinfo {author} {\bibfnamefont {G.}~\bibnamefont {Neubauer}},\ }\href@noop
  {} {\emph {\bibinfo {title} {Regularization of Inverse Problems}}},\
  Mathematics and Its Applications\ (\bibinfo  {publisher} {Springer},\
  \bibinfo {year} {1996})\BibitemShut {NoStop}%
\bibitem [{\citenamefont {Bengtsson}\ and\ \citenamefont
  {Zyczkowski}(2006)}]{Bengtsson2006May}%
  \BibitemOpen
  \bibfield  {author} {\bibinfo {author} {\bibfnamefont {I.}~\bibnamefont
  {Bengtsson}}\ and\ \bibinfo {author} {\bibfnamefont {K.}~\bibnamefont
  {Zyczkowski}},\ }\href {https://doi.org/10.1017/CBO9780511535048} {\emph
  {\bibinfo {title} {{Geometry of Quantum States: An Introduction to Quantum
  Entanglement}}}}\ (\bibinfo  {publisher} {Cambridge University Press},\
  \bibinfo {address} {Cambridge, England, UK},\ \bibinfo {year}
  {2006})\BibitemShut {NoStop}%
\bibitem [{\citenamefont {Vogel}(2006)}]{Vogel2006Aug}%
  \BibitemOpen
  \bibfield  {author} {\bibinfo {author} {\bibfnamefont {C.~R.}\ \bibnamefont
  {Vogel}},\ }\bibfield  {title} {\bibinfo {title} {{A Constrained Least
  Squares Regularization Method for Nonlinear III-Posed Problems}},\ }\href
  {https://epubs.siam.org/doi/10.1137/0328002} {\bibfield  {journal} {\bibinfo
  {journal} {SIAM J. Control Optim.}\ } (\bibinfo {year} {2006})}\BibitemShut
  {NoStop}%
\bibitem [{\citenamefont {Souopgui}\ \emph {et~al.}(2016)\citenamefont
  {Souopgui}, \citenamefont {Ngodock}, \citenamefont {Vidard},\ and\
  \citenamefont {Le~Dimet}}]{Souopgui2016Oct}%
  \BibitemOpen
  \bibfield  {author} {\bibinfo {author} {\bibfnamefont {I.}~\bibnamefont
  {Souopgui}}, \bibinfo {author} {\bibfnamefont {H.~E.}\ \bibnamefont
  {Ngodock}}, \bibinfo {author} {\bibfnamefont {A.}~\bibnamefont {Vidard}},\
  and\ \bibinfo {author} {\bibfnamefont {F.-X.}\ \bibnamefont {Le~Dimet}},\
  }\bibfield  {title} {\bibinfo {title} {{Incremental projection approach of
  regularization for inverse problems}},\ }\href
  {https://doi.org/10.1007/s00245-015-9315-3} {\bibfield  {journal} {\bibinfo
  {journal} {Appl. Math. Optim.}\ }\textbf {\bibinfo {volume} {74}},\ \bibinfo
  {pages} {303} (\bibinfo {year} {2016})}\BibitemShut {NoStop}%
\bibitem [{\citenamefont {Kirsch}(2011)}]{Kirsch2011-zx}%
  \BibitemOpen
  \bibfield  {author} {\bibinfo {author} {\bibfnamefont {A.}~\bibnamefont
  {Kirsch}},\ }\href@noop {} {\emph {\bibinfo {title} {An introduction to the
  mathematical theory of inverse problems}}},\ \bibinfo {edition} {2nd}\ ed.,\
  Applied mathematical sciences\ (\bibinfo  {publisher} {Springer},\ \bibinfo
  {year} {2011})\BibitemShut {NoStop}%
\bibitem [{\citenamefont {Diamond}\ and\ \citenamefont
  {Boyd}(2016)}]{Diamond2016}%
  \BibitemOpen
  \bibfield  {author} {\bibinfo {author} {\bibfnamefont {S.}~\bibnamefont
  {Diamond}}\ and\ \bibinfo {author} {\bibfnamefont {S.}~\bibnamefont {Boyd}},\
  }\bibfield  {title} {\bibinfo {title} {{CVXPY: A Python-Embedded Modeling
  Language for Convex Optimization}},\ }\href
  {https://pubmed.ncbi.nlm.nih.gov/27375369} {\bibfield  {journal} {\bibinfo
  {journal} {Journal of machine learning research : JMLR}\ }\textbf {\bibinfo
  {volume} {17}},\ \bibinfo {pages} {83} (\bibinfo {year} {2016})}\BibitemShut
  {NoStop}%
\bibitem [{\citenamefont {Agrawal}\ \emph {et~al.}(2018)\citenamefont
  {Agrawal}, \citenamefont {Verschueren}, \citenamefont {Diamond},\ and\
  \citenamefont {Boyd}}]{Agrawal2018Jan}%
  \BibitemOpen
  \bibfield  {author} {\bibinfo {author} {\bibfnamefont {A.}~\bibnamefont
  {Agrawal}}, \bibinfo {author} {\bibfnamefont {R.}~\bibnamefont
  {Verschueren}}, \bibinfo {author} {\bibfnamefont {S.}~\bibnamefont
  {Diamond}},\ and\ \bibinfo {author} {\bibfnamefont {S.}~\bibnamefont
  {Boyd}},\ }\bibfield  {title} {\bibinfo {title} {{A rewriting system for
  convex optimization problems}},\ }\href
  {https://doi.org/10.1080/23307706.2017.1397554} {\bibfield  {journal}
  {\bibinfo  {journal} {Journal of Control and Decision}\ }\textbf {\bibinfo
  {volume} {5}},\ \bibinfo {pages} {42} (\bibinfo {year} {2018})}\BibitemShut
  {NoStop}%
\bibitem [{Bib(2021)}]{BibEntry2021Jun}%
  \BibitemOpen
  \href {https://www.cvxpy.org/index.html} {\bibinfo {title} {{CVXPY website}}}
  (\bibinfo {year} {2021}),\ \bibinfo {note} {[Online; accessed 9. Jul.
  2021]}\BibitemShut {NoStop}%
\bibitem [{\citenamefont {Leonhardt}\ and\ \citenamefont
  {Paul}(1995)}]{Leonhardt1995Jan}%
  \BibitemOpen
  \bibfield  {author} {\bibinfo {author} {\bibfnamefont {U.}~\bibnamefont
  {Leonhardt}}\ and\ \bibinfo {author} {\bibfnamefont {H.}~\bibnamefont
  {Paul}},\ }\bibfield  {title} {\bibinfo {title} {{Measuring the quantum state
  of light}},\ }\href {https://doi.org/10.1016/0079-6727(94)00007-L} {\bibfield
   {journal} {\bibinfo  {journal} {Prog. Quantum Electron.}\ }\textbf {\bibinfo
  {volume} {19}},\ \bibinfo {pages} {89} (\bibinfo {year} {1995})}\BibitemShut
  {NoStop}%
\bibitem [{\citenamefont {Sych}\ \emph {et~al.}(2012)\citenamefont {Sych},
  \citenamefont
  {{\ifmmode\check{R}\else\v{R}\fi}eh{\ifmmode\acute{a}\else\'{a}\fi}{\ifmmode\check{c}\else\v{c}\fi}ek},
  \citenamefont {Hradil}, \citenamefont {Leuchs},\ and\ \citenamefont
  {S{\ifmmode\acute{a}\else\'{a}\fi}nchez-Soto}}]{Sych2012Nov}%
  \BibitemOpen
  \bibfield  {author} {\bibinfo {author} {\bibfnamefont {D.}~\bibnamefont
  {Sych}}, \bibinfo {author} {\bibfnamefont {J.}~\bibnamefont
  {{\ifmmode\check{R}\else\v{R}\fi}eh{\ifmmode\acute{a}\else\'{a}\fi}{\ifmmode\check{c}\else\v{c}\fi}ek}},
  \bibinfo {author} {\bibfnamefont {Z.}~\bibnamefont {Hradil}}, \bibinfo
  {author} {\bibfnamefont {G.}~\bibnamefont {Leuchs}},\ and\ \bibinfo {author}
  {\bibfnamefont {L.~L.}\ \bibnamefont
  {S{\ifmmode\acute{a}\else\'{a}\fi}nchez-Soto}},\ }\bibfield  {title}
  {\bibinfo {title} {{Informational completeness of continuous-variable
  measurements}},\ }\href {https://doi.org/10.1103/PhysRevA.86.052123}
  {\bibfield  {journal} {\bibinfo  {journal} {Phys. Rev. A}\ }\textbf {\bibinfo
  {volume} {86}},\ \bibinfo {pages} {052123} (\bibinfo {year}
  {2012})}\BibitemShut {NoStop}%
\bibitem [{\citenamefont {Strandberg}\ \emph {et~al.}(2019)\citenamefont
  {Strandberg}, \citenamefont {Lu}, \citenamefont
  {Quijandr{\ifmmode\acute{\imath}\else\'{\i}\fi}a},\ and\ \citenamefont
  {Johansson}}]{Strandberg2019Dec}%
  \BibitemOpen
  \bibfield  {author} {\bibinfo {author} {\bibfnamefont {I.}~\bibnamefont
  {Strandberg}}, \bibinfo {author} {\bibfnamefont {Y.}~\bibnamefont {Lu}},
  \bibinfo {author} {\bibfnamefont {F.}~\bibnamefont
  {Quijandr{\ifmmode\acute{\imath}\else\'{\i}\fi}a}},\ and\ \bibinfo {author}
  {\bibfnamefont {G.}~\bibnamefont {Johansson}},\ }\bibfield  {title} {\bibinfo
  {title} {{Numerical study of Wigner negativity in one-dimensional
  steady-state resonance fluorescence}},\ }\href
  {https://doi.org/10.1103/PhysRevA.100.063808} {\bibfield  {journal} {\bibinfo
   {journal} {Phys. Rev. A}\ }\textbf {\bibinfo {volume} {100}},\ \bibinfo
  {pages} {063808} (\bibinfo {year} {2019})}\BibitemShut {NoStop}%
\bibitem [{\citenamefont {Barnett}\ and\ \citenamefont
  {Radmore}(2002)}]{barnett2002methods}%
  \BibitemOpen
  \bibfield  {author} {\bibinfo {author} {\bibfnamefont {S.}~\bibnamefont
  {Barnett}}\ and\ \bibinfo {author} {\bibfnamefont {P.}~\bibnamefont
  {Radmore}},\ }\href@noop {} {\emph {\bibinfo {title} {Methods in Theoretical
  Quantum Optics}}},\ Oxford Series in Optical and Imaging Sciences\ (\bibinfo
  {publisher} {Clarendon Press},\ \bibinfo {year} {2002})\ p.~\bibinfo {pages}
  {37}\BibitemShut {NoStop}%
\bibitem [{\citenamefont {Leonhardt}(1997)}]{Leonhardt1997-tn}%
  \BibitemOpen
  \bibfield  {author} {\bibinfo {author} {\bibfnamefont {U.}~\bibnamefont
  {Leonhardt}},\ }\href@noop {} {\emph {\bibinfo {title} {Measuring the quantum
  state of light}}},\ Cambridge studies in modern optics\ (\bibinfo
  {publisher} {Cambridge University Press},\ \bibinfo {year}
  {1997})\BibitemShut {NoStop}%
\bibitem [{\citenamefont {Wiseman}\ and\ \citenamefont
  {Milburn}(1993)}]{Wiseman1993Jan}%
  \BibitemOpen
  \bibfield  {author} {\bibinfo {author} {\bibfnamefont {H.~M.}\ \bibnamefont
  {Wiseman}}\ and\ \bibinfo {author} {\bibfnamefont {G.~J.}\ \bibnamefont
  {Milburn}},\ }\bibfield  {title} {\bibinfo {title} {{Quantum theory of
  field-quadrature measurements}},\ }\href
  {https://doi.org/10.1103/PhysRevA.47.642} {\bibfield  {journal} {\bibinfo
  {journal} {Phys. Rev. A}\ }\textbf {\bibinfo {volume} {47}},\ \bibinfo
  {pages} {642} (\bibinfo {year} {1993})}\BibitemShut {NoStop}%
\bibitem [{\citenamefont {Strandberg}(2022)}]{code}%
  \BibitemOpen
  \bibfield  {author} {\bibinfo {author} {\bibfnamefont {I.}~\bibnamefont
  {Strandberg}},\ }\href {https://doi.org/10.5281/zenodo.6912651} {\bibinfo
  {title} {Program code: cvx-tomography}} (\bibinfo {year} {2022}),\ \bibinfo
  {note} {also on
  \href{https://github.com/ingstra/cvx-tomography}{GitHub}}\BibitemShut
  {NoStop}%
\bibitem [{\citenamefont {Rundle}\ and\ \citenamefont
  {Everitt}(2021)}]{Rundle2021Jun}%
  \BibitemOpen
  \bibfield  {author} {\bibinfo {author} {\bibfnamefont {R.~P.}\ \bibnamefont
  {Rundle}}\ and\ \bibinfo {author} {\bibfnamefont {M.~J.}\ \bibnamefont
  {Everitt}},\ }\bibfield  {title} {\bibinfo {title} {{Overview of the Phase
  Space Formulation of Quantum Mechanics with Application to Quantum
  Technologies}},\ }\href {https://doi.org/10.1002/qute.202100016} {\bibfield
  {journal} {\bibinfo  {journal} {Adv. Quantum Technol.}\ }\textbf {\bibinfo
  {volume} {4}},\ \bibinfo {pages} {2100016} (\bibinfo {year}
  {2021})}\BibitemShut {NoStop}%
\bibitem [{\citenamefont {Mirrahimi}\ \emph {et~al.}(2014)\citenamefont
  {Mirrahimi}, \citenamefont {Leghtas}, \citenamefont {Albert}, \citenamefont
  {Touzard}, \citenamefont {Schoelkopf}, \citenamefont {Jiang},\ and\
  \citenamefont {Devoret}}]{Mirrahimi2014Apr}%
  \BibitemOpen
  \bibfield  {author} {\bibinfo {author} {\bibfnamefont {M.}~\bibnamefont
  {Mirrahimi}}, \bibinfo {author} {\bibfnamefont {Z.}~\bibnamefont {Leghtas}},
  \bibinfo {author} {\bibfnamefont {V.~V.}\ \bibnamefont {Albert}}, \bibinfo
  {author} {\bibfnamefont {S.}~\bibnamefont {Touzard}}, \bibinfo {author}
  {\bibfnamefont {R.~J.}\ \bibnamefont {Schoelkopf}}, \bibinfo {author}
  {\bibfnamefont {L.}~\bibnamefont {Jiang}},\ and\ \bibinfo {author}
  {\bibfnamefont {M.~H.}\ \bibnamefont {Devoret}},\ }\bibfield  {title}
  {\bibinfo {title} {{Dynamically protected cat-qubits: a new paradigm for
  universal quantum computation}},\ }\href
  {https://doi.org/10.1088/1367-2630/16/4/045014} {\bibfield  {journal}
  {\bibinfo  {journal} {New J. Phys.}\ }\textbf {\bibinfo {volume} {16}},\
  \bibinfo {pages} {045014} (\bibinfo {year} {2014})}\BibitemShut {NoStop}%
\bibitem [{\citenamefont {Clerk}\ \emph {et~al.}(2010)\citenamefont {Clerk},
  \citenamefont {Devoret}, \citenamefont {Girvin}, \citenamefont {Marquardt},\
  and\ \citenamefont {Schoelkopf}}]{Clerk2010Apr}%
  \BibitemOpen
  \bibfield  {author} {\bibinfo {author} {\bibfnamefont {A.~A.}\ \bibnamefont
  {Clerk}}, \bibinfo {author} {\bibfnamefont {M.~H.}\ \bibnamefont {Devoret}},
  \bibinfo {author} {\bibfnamefont {S.~M.}\ \bibnamefont {Girvin}}, \bibinfo
  {author} {\bibfnamefont {F.}~\bibnamefont {Marquardt}},\ and\ \bibinfo
  {author} {\bibfnamefont {R.~J.}\ \bibnamefont {Schoelkopf}},\ }\bibfield
  {title} {\bibinfo {title} {{Introduction to quantum noise, measurement, and
  amplification}},\ }\href {https://doi.org/10.1103/RevModPhys.82.1155}
  {\bibfield  {journal} {\bibinfo  {journal} {Rev. Mod. Phys.}\ }\textbf
  {\bibinfo {volume} {82}},\ \bibinfo {pages} {1155} (\bibinfo {year}
  {2010})}\BibitemShut {NoStop}%
\bibitem [{\citenamefont {Kim}(1997)}]{Kim1997Oct}%
  \BibitemOpen
  \bibfield  {author} {\bibinfo {author} {\bibfnamefont {M.~S.}\ \bibnamefont
  {Kim}},\ }\bibfield  {title} {\bibinfo {title} {{Quasiprobability functions
  measured by photon statistics of amplified signal fields}},\ }\href
  {https://doi.org/10.1103/PhysRevA.56.3175} {\bibfield  {journal} {\bibinfo
  {journal} {Phys. Rev. A}\ }\textbf {\bibinfo {volume} {56}},\ \bibinfo
  {pages} {3175} (\bibinfo {year} {1997})}\BibitemShut {NoStop}%
\bibitem [{\citenamefont {Cahill}\ and\ \citenamefont
  {Glauber}(1969)}]{Cahill1969Jan}%
  \BibitemOpen
  \bibfield  {author} {\bibinfo {author} {\bibfnamefont {K.~E.}\ \bibnamefont
  {Cahill}}\ and\ \bibinfo {author} {\bibfnamefont {R.~J.}\ \bibnamefont
  {Glauber}},\ }\bibfield  {title} {\bibinfo {title} {{Density Operators and
  Quasiprobability Distributions}},\ }\href
  {https://doi.org/10.1103/PhysRev.177.1882} {\bibfield  {journal} {\bibinfo
  {journal} {Phys. Rev.}\ }\textbf {\bibinfo {volume} {177}},\ \bibinfo {pages}
  {1882} (\bibinfo {year} {1969})}\BibitemShut {NoStop}%
\bibitem [{\citenamefont {Glauber}(1963)}]{Glauber1963Sep}%
  \BibitemOpen
  \bibfield  {author} {\bibinfo {author} {\bibfnamefont {R.~J.}\ \bibnamefont
  {Glauber}},\ }\bibfield  {title} {\bibinfo {title} {{Coherent and Incoherent
  States of the Radiation Field}},\ }\href
  {https://doi.org/10.1103/PhysRev.131.2766} {\bibfield  {journal} {\bibinfo
  {journal} {Phys. Rev.}\ }\textbf {\bibinfo {volume} {131}},\ \bibinfo {pages}
  {2766} (\bibinfo {year} {1963})}\BibitemShut {NoStop}%
\bibitem [{\citenamefont {Eichler}(2013)}]{Eichler2013}%
  \BibitemOpen
  \bibfield  {author} {\bibinfo {author} {\bibfnamefont {C.}~\bibnamefont
  {Eichler}},\ }\emph {\bibinfo {title} {{Experimental characterization of
  quantum microwave radiation and its entanglement with a superconducting
  qubit}}},\ \href {https://doi.org/10.3929/ethz-a-009771047} {Ph.D. thesis},\
  \bibinfo  {school} {ETH}, \bibinfo {address}
  {Z{\ifmmode\ddot{u}\else\"{u}\fi}rich, Switzerland} (\bibinfo {year}
  {2013})\BibitemShut {NoStop}%
\bibitem [{\citenamefont {Ahmed}(2021)}]{shahnawaz_ahmed_2021_5105470}%
  \BibitemOpen
  \bibfield  {author} {\bibinfo {author} {\bibfnamefont {S.}~\bibnamefont
  {Ahmed}},\ }\href {https://doi.org/10.5281/zenodo.5105470} {\bibinfo {title}
  {Code for quantum state tomography with conditional generative adversarial
  networks}} (\bibinfo {year} {2021}),\ \bibinfo {note}
  {\url{https://doi.org/10.5281/zenodo.5105470}}\BibitemShut {NoStop}%
\bibitem [{\citenamefont {Lu}\ \emph {et~al.}(2021{\natexlab{a}})\citenamefont
  {Lu}, \citenamefont {Strandberg}, \citenamefont
  {Quijandr{\ifmmode\acute{\imath}\else\'{\i}\fi}a}, \citenamefont {Johansson},
  \citenamefont {Gasparinetti},\ and\ \citenamefont {Delsing}}]{Lu2021Jun}%
  \BibitemOpen
  \bibfield  {author} {\bibinfo {author} {\bibfnamefont {Y.}~\bibnamefont
  {Lu}}, \bibinfo {author} {\bibfnamefont {I.}~\bibnamefont {Strandberg}},
  \bibinfo {author} {\bibfnamefont {F.}~\bibnamefont
  {Quijandr{\ifmmode\acute{\imath}\else\'{\i}\fi}a}}, \bibinfo {author}
  {\bibfnamefont {G.}~\bibnamefont {Johansson}}, \bibinfo {author}
  {\bibfnamefont {S.}~\bibnamefont {Gasparinetti}},\ and\ \bibinfo {author}
  {\bibfnamefont {P.}~\bibnamefont {Delsing}},\ }\bibfield  {title} {\bibinfo
  {title} {{Propagating Wigner-Negative States Generated from the Steady-State
  Emission of a Superconducting Qubit}},\ }\href
  {https://doi.org/10.1103/PhysRevLett.126.253602} {\bibfield  {journal}
  {\bibinfo  {journal} {Phys. Rev. Lett.}\ }\textbf {\bibinfo {volume} {126}},\
  \bibinfo {pages} {253602} (\bibinfo {year} {2021}{\natexlab{a}})}\BibitemShut
  {NoStop}%
\bibitem [{\citenamefont {Eichler}\ \emph {et~al.}(2011)\citenamefont
  {Eichler}, \citenamefont {Bozyigit}, \citenamefont {Lang}, \citenamefont
  {Steffen}, \citenamefont {Fink},\ and\ \citenamefont
  {Wallraff}}]{Eichler2011Jun}%
  \BibitemOpen
  \bibfield  {author} {\bibinfo {author} {\bibfnamefont {C.}~\bibnamefont
  {Eichler}}, \bibinfo {author} {\bibfnamefont {D.}~\bibnamefont {Bozyigit}},
  \bibinfo {author} {\bibfnamefont {C.}~\bibnamefont {Lang}}, \bibinfo {author}
  {\bibfnamefont {L.}~\bibnamefont {Steffen}}, \bibinfo {author} {\bibfnamefont
  {J.}~\bibnamefont {Fink}},\ and\ \bibinfo {author} {\bibfnamefont
  {A.}~\bibnamefont {Wallraff}},\ }\bibfield  {title} {\bibinfo {title}
  {{Experimental State Tomography of Itinerant Single Microwave Photons}},\
  }\href {https://doi.org/10.1103/PhysRevLett.106.220503} {\bibfield  {journal}
  {\bibinfo  {journal} {Phys. Rev. Lett.}\ }\textbf {\bibinfo {volume} {106}},\
  \bibinfo {pages} {220503} (\bibinfo {year} {2011})}\BibitemShut {NoStop}%
\bibitem [{\citenamefont {Eichler}\ \emph {et~al.}(2012)\citenamefont
  {Eichler}, \citenamefont {Bozyigit},\ and\ \citenamefont
  {Wallraff}}]{Eichler2012Sep}%
  \BibitemOpen
  \bibfield  {author} {\bibinfo {author} {\bibfnamefont {C.}~\bibnamefont
  {Eichler}}, \bibinfo {author} {\bibfnamefont {D.}~\bibnamefont {Bozyigit}},\
  and\ \bibinfo {author} {\bibfnamefont {A.}~\bibnamefont {Wallraff}},\
  }\bibfield  {title} {\bibinfo {title} {{Characterizing quantum microwave
  radiation and its entanglement with superconducting qubits using linear
  detectors}},\ }\href {https://doi.org/10.1103/PhysRevA.86.032106} {\bibfield
  {journal} {\bibinfo  {journal} {Phys. Rev. A}\ }\textbf {\bibinfo {volume}
  {86}},\ \bibinfo {pages} {032106} (\bibinfo {year} {2012})}\BibitemShut
  {NoStop}%
\bibitem [{\citenamefont {Lu}\ \emph {et~al.}(2021{\natexlab{b}})\citenamefont
  {Lu}, \citenamefont {Bengtsson}, \citenamefont {Burnett}, \citenamefont
  {Suri}, \citenamefont {Sathyamoorthy}, \citenamefont {Nilsson}, \citenamefont
  {Scigliuzzo}, \citenamefont {Bylander}, \citenamefont {Johansson},\ and\
  \citenamefont {Delsing}}]{Lu2021Sep}%
  \BibitemOpen
  \bibfield  {author} {\bibinfo {author} {\bibfnamefont {Y.}~\bibnamefont
  {Lu}}, \bibinfo {author} {\bibfnamefont {A.}~\bibnamefont {Bengtsson}},
  \bibinfo {author} {\bibfnamefont {J.~J.}\ \bibnamefont {Burnett}}, \bibinfo
  {author} {\bibfnamefont {B.}~\bibnamefont {Suri}}, \bibinfo {author}
  {\bibfnamefont {S.~R.}\ \bibnamefont {Sathyamoorthy}}, \bibinfo {author}
  {\bibfnamefont {H.~R.}\ \bibnamefont {Nilsson}}, \bibinfo {author}
  {\bibfnamefont {M.}~\bibnamefont {Scigliuzzo}}, \bibinfo {author}
  {\bibfnamefont {J.}~\bibnamefont {Bylander}}, \bibinfo {author}
  {\bibfnamefont {G.}~\bibnamefont {Johansson}},\ and\ \bibinfo {author}
  {\bibfnamefont {P.}~\bibnamefont {Delsing}},\ }\bibfield  {title} {\bibinfo
  {title} {{Quantum efficiency, purity and stability of a tunable, narrowband
  microwave single-photon source - npj Quantum Information}},\ }\href
  {https://doi.org/10.1038/s41534-021-00480-5} {\bibfield  {journal} {\bibinfo
  {journal} {npj Quantum Inf.}\ }\textbf {\bibinfo {volume} {7}},\ \bibinfo
  {pages} {1} (\bibinfo {year} {2021}{\natexlab{b}})}\BibitemShut {NoStop}%
\bibitem [{\citenamefont {Royer}(1977)}]{Royer1977Feb}%
  \BibitemOpen
  \bibfield  {author} {\bibinfo {author} {\bibfnamefont {A.}~\bibnamefont
  {Royer}},\ }\bibfield  {title} {\bibinfo {title} {{Wigner function as the
  expectation value of a parity operator}},\ }\href
  {https://doi.org/10.1103/PhysRevA.15.449} {\bibfield  {journal} {\bibinfo
  {journal} {Phys. Rev. A}\ }\textbf {\bibinfo {volume} {15}},\ \bibinfo
  {pages} {449} (\bibinfo {year} {1977})}\BibitemShut {NoStop}%
\bibitem [{\citenamefont {Banaszek}\ \emph
  {et~al.}(1999{\natexlab{b}})\citenamefont {Banaszek}, \citenamefont
  {Radzewicz}, \citenamefont {W{\ifmmode\acute{o}\else\'{o}\fi}dkiewicz},\ and\
  \citenamefont {Krasi{\ifmmode\acute{n}\else\'{n}\fi}ski}}]{Banaszek1999Jul}%
  \BibitemOpen
  \bibfield  {author} {\bibinfo {author} {\bibfnamefont {K.}~\bibnamefont
  {Banaszek}}, \bibinfo {author} {\bibfnamefont {C.}~\bibnamefont {Radzewicz}},
  \bibinfo {author} {\bibfnamefont {K.}~\bibnamefont
  {W{\ifmmode\acute{o}\else\'{o}\fi}dkiewicz}},\ and\ \bibinfo {author}
  {\bibfnamefont {J.~S.}\ \bibnamefont
  {Krasi{\ifmmode\acute{n}\else\'{n}\fi}ski}},\ }\bibfield  {title} {\bibinfo
  {title} {{Direct measurement of the Wigner function by photon counting}},\
  }\href {https://doi.org/10.1103/PhysRevA.60.674} {\bibfield  {journal}
  {\bibinfo  {journal} {Phys. Rev. A}\ }\textbf {\bibinfo {volume} {60}},\
  \bibinfo {pages} {674} (\bibinfo {year} {1999}{\natexlab{b}})}\BibitemShut
  {NoStop}%
\bibitem [{\citenamefont {Kudra}\ \emph {et~al.}(2021)\citenamefont {Kudra},
  \citenamefont {Kervinen}, \citenamefont {Strandberg}, \citenamefont {Ahmed},
  \citenamefont {Scigliuzzo}, \citenamefont {Osman}, \citenamefont {Lozano},
  \citenamefont {Ferrini}, \citenamefont {Bylander}, \citenamefont {Kockum},
  \citenamefont {Quijandr{\ifmmode\acute{\imath}\else\'{\i}\fi}a},
  \citenamefont {Delsing},\ and\ \citenamefont {Gasparinetti}}]{Kudra2021Nov}%
  \BibitemOpen
  \bibfield  {author} {\bibinfo {author} {\bibfnamefont {M.}~\bibnamefont
  {Kudra}}, \bibinfo {author} {\bibfnamefont {M.}~\bibnamefont {Kervinen}},
  \bibinfo {author} {\bibfnamefont {I.}~\bibnamefont {Strandberg}}, \bibinfo
  {author} {\bibfnamefont {S.}~\bibnamefont {Ahmed}}, \bibinfo {author}
  {\bibfnamefont {M.}~\bibnamefont {Scigliuzzo}}, \bibinfo {author}
  {\bibfnamefont {A.}~\bibnamefont {Osman}}, \bibinfo {author} {\bibfnamefont
  {D.~P.}\ \bibnamefont {Lozano}}, \bibinfo {author} {\bibfnamefont
  {G.}~\bibnamefont {Ferrini}}, \bibinfo {author} {\bibfnamefont
  {J.}~\bibnamefont {Bylander}}, \bibinfo {author} {\bibfnamefont {A.~F.}\
  \bibnamefont {Kockum}}, \bibinfo {author} {\bibfnamefont {F.}~\bibnamefont
  {Quijandr{\ifmmode\acute{\imath}\else\'{\i}\fi}a}}, \bibinfo {author}
  {\bibfnamefont {P.}~\bibnamefont {Delsing}},\ and\ \bibinfo {author}
  {\bibfnamefont {S.}~\bibnamefont {Gasparinetti}},\ }\bibfield  {title}
  {\bibinfo {title} {{Robust preparation of Wigner-negative states with
  optimized SNAP-displacement sequences}},\ }\href
  {https://arxiv.org/abs/2111.07965v1} {\bibfield  {journal} {\bibinfo
  {journal} {arXiv}\ } (\bibinfo {year} {2021})},\ \Eprint
  {https://arxiv.org/abs/2111.07965} {2111.07965} \BibitemShut {NoStop}%
\bibitem [{\citenamefont {Leonhardt}\ and\ \citenamefont
  {Paul}(1993)}]{Leonhardt1993Dec}%
  \BibitemOpen
  \bibfield  {author} {\bibinfo {author} {\bibfnamefont {U.}~\bibnamefont
  {Leonhardt}}\ and\ \bibinfo {author} {\bibfnamefont {H.}~\bibnamefont
  {Paul}},\ }\bibfield  {title} {\bibinfo {title} {{Realistic optical homodyne
  measurements and quasiprobability distributions}},\ }\href
  {https://doi.org/10.1103/PhysRevA.48.4598} {\bibfield  {journal} {\bibinfo
  {journal} {Phys. Rev. A}\ }\textbf {\bibinfo {volume} {48}},\ \bibinfo
  {pages} {4598} (\bibinfo {year} {1993})}\BibitemShut {NoStop}%
\bibitem [{\citenamefont {Kubo}(1966)}]{Kubo1966Jan}%
  \BibitemOpen
  \bibfield  {author} {\bibinfo {author} {\bibfnamefont {R.}~\bibnamefont
  {Kubo}},\ }\bibfield  {title} {\bibinfo {title} {{The fluctuation-dissipation
  theorem}},\ }\href {https://doi.org/10.1088/0034-4885/29/1/306} {\bibfield
  {journal} {\bibinfo  {journal} {Rep. Prog. Phys.}\ }\textbf {\bibinfo
  {volume} {29}},\ \bibinfo {pages} {255} (\bibinfo {year} {1966})}\BibitemShut
  {NoStop}%
\bibitem [{\citenamefont {O'Donoghue}\ \emph {et~al.}(2016)\citenamefont
  {O'Donoghue}, \citenamefont {Chu}, \citenamefont {Parikh},\ and\
  \citenamefont {Boyd}}]{ocpb:16}%
  \BibitemOpen
  \bibfield  {author} {\bibinfo {author} {\bibfnamefont {B.}~\bibnamefont
  {O'Donoghue}}, \bibinfo {author} {\bibfnamefont {E.}~\bibnamefont {Chu}},
  \bibinfo {author} {\bibfnamefont {N.}~\bibnamefont {Parikh}},\ and\ \bibinfo
  {author} {\bibfnamefont {S.}~\bibnamefont {Boyd}},\ }\bibfield  {title}
  {\bibinfo {title} {Conic optimization via operator splitting and homogeneous
  self-dual embedding},\ }\href {http://stanford.edu/~boyd/papers/scs.html}
  {\bibfield  {journal} {\bibinfo  {journal} {Journal of Optimization Theory
  and Applications}\ }\textbf {\bibinfo {volume} {169}},\ \bibinfo {pages}
  {1042} (\bibinfo {year} {2016})}\BibitemShut {NoStop}%
\bibitem [{\citenamefont {O'Donoghue}\ \emph {et~al.}(2021)\citenamefont
  {O'Donoghue}, \citenamefont {Chu}, \citenamefont {Parikh},\ and\
  \citenamefont {Boyd}}]{scs}%
  \BibitemOpen
  \bibfield  {author} {\bibinfo {author} {\bibfnamefont {B.}~\bibnamefont
  {O'Donoghue}}, \bibinfo {author} {\bibfnamefont {E.}~\bibnamefont {Chu}},
  \bibinfo {author} {\bibfnamefont {N.}~\bibnamefont {Parikh}},\ and\ \bibinfo
  {author} {\bibfnamefont {S.}~\bibnamefont {Boyd}},\ }\href@noop {} {\bibinfo
  {title} {{SCS}: Splitting conic solver, version 3.2.1}},\ \bibinfo
  {howpublished} {\url{https://github.com/cvxgrp/scs}} (\bibinfo {year}
  {2021})\BibitemShut {NoStop}%
\bibitem [{\citenamefont {Rontsis}\ \emph {et~al.}(2022)\citenamefont
  {Rontsis}, \citenamefont {Goulart},\ and\ \citenamefont
  {Nakatsukasa}}]{Rontsis2022Jan}%
  \BibitemOpen
  \bibfield  {author} {\bibinfo {author} {\bibfnamefont {N.}~\bibnamefont
  {Rontsis}}, \bibinfo {author} {\bibfnamefont {P.}~\bibnamefont {Goulart}},\
  and\ \bibinfo {author} {\bibfnamefont {Y.}~\bibnamefont {Nakatsukasa}},\
  }\bibfield  {title} {\bibinfo {title} {{Efficient Semidefinite Programming
  with Approximate ADMM}},\ }\href {https://doi.org/10.1007/s10957-021-01971-3}
  {\bibfield  {journal} {\bibinfo  {journal} {J. Optim. Theory Appl.}\ }\textbf
  {\bibinfo {volume} {192}},\ \bibinfo {pages} {292} (\bibinfo {year}
  {2022})}\BibitemShut {NoStop}%
\bibitem [{\citenamefont {Allaire}\ and\ \citenamefont
  {Kaber}(2008)}]{Allaire2008}%
  \BibitemOpen
  \bibfield  {author} {\bibinfo {author} {\bibfnamefont {G.}~\bibnamefont
  {Allaire}}\ and\ \bibinfo {author} {\bibfnamefont {S.~M.}\ \bibnamefont
  {Kaber}},\ }\href {https://link.springer.com/book/10.1007/978-0-387-68918-0}
  {\emph {\bibinfo {title} {{Numerical Linear Algebra}}}}\ (\bibinfo
  {publisher} {Springer},\ \bibinfo {address} {New York, NY, USA},\ \bibinfo
  {year} {2008})\BibitemShut {NoStop}%
\end{thebibliography}%

\end{document}